\documentclass[]{article}

\usepackage{graphicx,amsmath,amsthm,amssymb,graphicx,titlesec,blindtext,color,mathrsfs}
\usepackage[english]{babel}
\usepackage{cite}

\def\R{\mathbb{R}}

\theoremstyle{plain}
\newtheorem{theorem}{Theorem}
\newtheorem{proposition}[theorem]{Proposition}

\theoremstyle{definition}

\newcommand{\hsp}{\hspace{20pt}}
\titleformat{\chapter}[hang]{\Huge\bfseries}{\thechapter\hsp$\mathbf{|}$\hsp}{0pt}{\Huge\bfseries}

\title{A fractional Brownian - Hawkes model for \\the Italian electricity spot market: \\ estimation and forecasting}
\author{Luca M.\ Giordano, Daniela Morale\\{\small{Department of Matemathics, Univeristy of Milano}}}
 
\begin{document}
	\maketitle
\begin{abstract}

We propose a model for the description and the forecast of the gross
prices of electricity in the liberalized Italian energy market  via an additive two-factor model driven by both a Hawkes   and a fractional Brownian processes.  We discuss the seasonality, the identification of spikes and the estimates of the Hurst coefficient. After the calibration and the validation of the model, we discuss its forecasting performance via a class of adequate evaluation metrics.

\end{abstract}

\section{Mathematical modelling of electricity markets}

\medskip  

This paper presents a modelling and computational work,
related to the description and forecasting of prices in the Italian wholesale electricity market. We discuss the choice of the particular model, above all   the representation of the source of randomness, related to the baseline evolution of prices via a fractional Brownian motion and to the spiky behaviour via a Hawkes process. 
The computational part of the work regards to problems of parameter estimation and dataset filtering. These are crucial steps in the pre-processing phase of the model. We  evaluate the performance of the model by means of the production of forecasts of future electricity prices, at different forecasting horizons with the  aim of evaluating more into details the
quality of the forecasts in the distributional sense, instead of giving a single prediction value. Adequate metrics are taken into consideration, as the Winkler score
and the Pinball loss function.

\medskip

In the last decades the electricity market has been liberalized in
a growing number of countries, especially in the European Union.
Liberalized markets have been introduced for example in Germany,
Italy, Spain, UK, as well as in all nordic countries. The introduction of
competitive markets has been reshaping the landscape of the
power sectors. Electricity price now undergoes to market rules, but it is very
peculiar.  Indeed, electricity is a non-storable commodity, hence the
need of having a particular organization in the market emerged.
This has usually resulted in the creation of a \textit{day-ahead
	market}: a market in which every day there are some auctions
regarding the delivery of energy at a fixed time of the following
day. The price of the electricity is determined by crossing
the supply curve and the demand curve, for the hour for which the
auction is taking place (see e.g.\ \cite{roncoroni_2015}). The
steepness of supply and demand curve can be regarded as the cause
of one of the main characteristics of the electricity price
market, i.e.\ \textit{shock formation} in the prices, which
is one of the most important aspects that distinguish the
electricity market from the other similar financial markets, and
also one of the most difficult to model. A \textit{shock}, or
\textit{spike} is a sudden large rise in the price followed by a
rapid drop to its regular level. 

\medskip

The distinction between spiky and  ``standard" behavior turns out
to be crucial in the modelling of the electricity price: the need
to obtain a good reproduction of the spikes in the
models represents one of the main differences between
other financial markets and the electricity one. Another
important difference with respect to other markets is the
seasonality that can be observed. It is mainly due to a clear
weekly periodicity, caused by the fluctuations in
consumptions during different days of the week
\cite{roncoroni_2015}. There is also a long-term seasonal effect on the prices, which appears over time lengths of approximately 3-4 months.

\medskip

There is a widespread mathematical literature about electricity spot market, both for the modelling  \cite{Barndorff-Nielsen_Benth_Veraart_2013,becker_clements_nur_2013,benth_libro_2008,benth_kallsen_meyer_brandis_2007,benth_Kiesel_Nazarova_2012,cartea_figueroa_2005,geman_roncoroni_2006,Gianfreda,sgarra_2018,schwartz_1997,schwartz_2002,Meyer-Brandis_2010,Meyer-Brandis_Tankov_2008,weron_review_2014,weron_2008,weron_2006}
and the calibration problems \cite{Janczura_spike_season_2013,Nowotarski_season_wavelets_2013,Weron_2018,weron_2006,weron_review_2014}.
Two of the first models for electricity price are due to Schwartz
\cite{schwartz_1997,schwartz_2002}. In \cite{schwartz_1997} the authors
introduced an Ornstein-Uhlenbeck model for the spot price dynamics
which included
a mean-reversion component, and later on, in  \cite{schwartz_2002}, a deterministic component describing the
seasonality was added. Since this works, a widespread literature has
been proposed in order to model the basic features of this
market, especially about the formation of spikes, which were not covered by the
aforementioned papers \cite{schwartz_1997,schwartz_2002}. An interesting review of the state of the art has been given by Weron in
\cite{weron_review_2014}. The interested reader may also refer to the huge amount of papers therein.
Weron proposes a classification of the models in five main
classes: i) \emph{the multi-agent  models}, which simulate the operation of a system of heterogeneous agents
interacting with each other, and build the price process by matching the demand and supply in the
market;
ii) \emph{the fundamental structural methods}, which describe the price
dynamics by modelling the impacts of important physical and
economic factors on the price of electricity; iii) \emph{the
reduced-form   stochastic  models}, which characterize
the statistical properties of electricity prices over time, with
the ultimate objective of derivatives evaluation and risk
management; iv) \emph{the statistical models}, which are either
direct applications of the statistical techniques of load
forecasting or power market implementations of econometric models;
v) \emph{the computational intelligence  techniques}, which combine
elements of learning, evolution and fuzziness to create approaches
that are capable of adapting to complex dynamic systems.

\bigskip

The model we present here mainly belongs to the class of the reduced-form model: 
the time evolution of the
spot price is described by an additive terms with a drift described by a deterministic
function which models  the long--term seasonality and two additive noises described as solution of two independent   stochastic differential equations. The   model has the following form
$$S(t)=f(t)+\sum_{i=1}^2 X_i(t), \quad
t\ge t_0 \in \mathbb R_+,$$
where $f$ is a deterministic function and the $X_i$, for $i=1,2$ are 
two stochastic processes,  responsible  of the standard fluctuations which give rise to the \textit{standard behavior}, i.e. the so called  \emph{base component}, and 
 the \textit{spiky behavior} of the price evolution, respectively. Both of them show   a mean reverting property.
 Different examples of these kind of models may be found for
example in
\cite{Meyer-Brandis_2010,Meyer-Brandis_Tankov_2008,weron_review_2014,sgarra_2018}.

\medskip

The novelty here is that we consider a fractional Brownian motion  as the driving noise of the base component instead of the usual standard Brownian motion. In particular, the process $X_1$ is a fractional Ornstein-Uhlenbeck process. The main modelling reason is that among the characteristics of the spot prices, one the presence of self-correlations in the price increments has to taken into account. The presence of this
feature suggests, when trying to model these kind of markets, to
modify the structure of the existing models to include the self-correlations. One of the possible choices that have been used in literature so far is to consider a fractionally integrated ARFIMA model, a generalisation of the classical ARIMA model, as it has been done in \cite{Gianfreda}, and in other cases reported in
the review \cite{weron_review_2014}. In particular, in \cite{Gianfreda} this has been done for the Italian electricity market. Here we go further than the pure statistical models.

\medskip

In literature there have been several attempts of using a fractional Brownian motion in
financial market modelling, even if not in the case of the electricity one. Its relatively simple nature, combined with its flexibility in
modelling data whose increments are self--correlated, gave rise to a growing number of models
involving fractional Brownian motion. Anyway, it was pointed out quite early   in \cite{rogers_1997} that a model
involving fBm would result in admitting the presence of some kind of arbitrage in the market.
More into details, in \cite{Cheridito_2002} the author proved that
there are strong arbitrage opportunities for the fractional models
of the form
\begin{equation}
\label{eq: models without arbitrage}
\begin{split}
X(t)= & \nu(t)+\sigma B^H(t) \\
X(t)= & \exp(\nu(t)+\sigma B^H(t)),\\
\end{split}
\end{equation}
where $\nu(t)$ is a measurable bounded deterministic function and
$B^H$ is a fractional Brownian motion of \textit{Hurst parameter}
$H\in(0,1)$. This arbitrage opportunities can be built provided
that we are allowed to use the typical set of admissible trading
strategies (see \cite{Cheridito_2002} for the complete
definitions). This set of admissible strategies in particular
allows to buy and sell the stock continuously in time, which is a
questionable assumption in many frameworks. In
\cite{Cheridito_2002} the author proved that for the models
\eqref{eq: models without arbitrage} the arbitrage
opportunities disappear, provided that we restrict the set of
strategies to the ones that impose an arbitrary (but fixed)
waiting time $h>0$ between a transaction and the following one. We recall that in
the present work we consider a fractional Ornstein--Uhlenbeck
process. We cannot use directly the results in
\cite{Cheridito_2002}, but the extension to this family of
processes should be straightforward and may be subject for future work. 

\medskip

A striking empirical feature of electricity spot prices is the presence of spikes, that can be described by a jump in the price process immediately followed by a fast reversion towards the mean.  It is  interesting to notice that in the case of the Italian electricity market   the presence of several jumps is shown, many of which appearing clustered over short time periods.
As a consequence, the second component $X_2$ is solution of a mean reverting processes driven by    a self-exciting Hawkes process, which is a jump process whose jumps frequency depends upon the previous history of the jump times. In particular, right after a jump has occurred, the probability of observing a subsequent jump is higher than usual. The interested reader may refer to
\cite{bacry_2015,hawkes_2018} fo an   excellent survey  on the introduction,
the relevant mathematical theory and overview of applications of
Hawkes processes in finance and for more recent financial
applications.

\medskip

We conclude the discussion on the model via some simulation results and the  problem of estimation  of the parameters of both the signals.
 
 \medskip

The second part of our paper is devoted to a complete computational study. We apply the model to the   study case of  the time series of the Italian
MGP, the data of the day-ahead market (see \cite{MPG}) from January 1, 2009   to December 31, 2017. The first two years are the sample
considered for the estimation and validation of the model. We   carry out the difficult task of separating the components of the raw prices into our main components (weekly component, long-term seasonal component, standard behaviour, spiky behaviour). Then we deal with the problem of the estimation the parameters of the model and we test the forecasting performance of our model on forecasting horizons from one to thirty days. The parameters are estimated in a \textit{rolling window} fashion, as summarized in \cite{Weron_2018}.   We construct prediction intervals (PI) and quantile forecasts (QF) and   evaluate them via a class of adequate evaluation metrics like the Winkler score and the Pinball loss function.
 
We conclude that the analysis shows some quantitative evidence that both the fractional Brownian motion and the Hawkes process are adequate to model the electricity price markets.

\medskip

The paper is organized as  follows. Section 2 is devoted to a brief presentation of both the fractional Brownian motion and the Hawkes processes with their main properties. We perform some numerical simulations for some  specific parameters. In Section 3 we discuss the method of estimation of the parameters, by validating the estimates by calculating the 5\% and the 95\% quantiles. Section 4 is dedicated to the study case. Preliminary analysis and a data  filtering process are implemented. A in sample estimation is carried out and then an out od sample forecasting is realized. The section ends with our conclusion.

\section{A model driven by a Hawkes  and a fractional Brownian processes}
\label{sec: model proposal}

The model we propose extends in different ways some relevant
models already available in the literature. In particular, we
consider a modification of the model proposed in
\cite{benth_kallsen_meyer_brandis_2007,Meyer-Brandis_Tankov_2008}
and then modified for example in \cite{sgarra_2018},  by including
some self--exciting features,  via Hawkes-type processes and correlation of the increments via a fractional Brownian motion.

\subsection{The equations}

We adopt an arithmetic model as in
\cite{Barndorff-Nielsen_Benth_Veraart_2013,benth_Kiesel_Nazarova_2012,benth_kallsen_meyer_brandis_2007,sgarra_2018}
in which the power price
dynamics is assumed to be the sum of several factors. We suppose
that the spot price process  $ S =\{S(t), t\in \mathbb R_+ \}$
evolves according to the following dynamics
\begin{equation}
\label{eq: model basic}
S(t)=f(t)+ {X}(t).
\end{equation}
The function $f(t)$ describes the deterministic trend of the
evolution, while the process   $ {X}=\{ {X}(t), t\in \mathbb
R_+\}$ describes the stochastic part. The latter is  a
superposition of two factors: $X_1$, known in literature as the
\textit{base component}, which is continuous a.e.\ and aims to
model the standard behavior of the electricity price, and $X_2$
which is the \textit{jump component}, describing the spiky
behavior of the electricity prices, overlapped to the base signal.
This means that, for any $t\in \mathbb R_+$,
\begin{equation}{X}(t)=X_1(t)+X_2(t).
\label{eq:X}
\end{equation}

The main reason to consider two additive signal driven by the two different noises is relating of our assumption of   mean-reversion:
despite the possible noise, the price tends to fluctuate around a specific level.
In particular, our starting assumption is that both in the
base and spiky regimes   prices tend to  revert towards their own  mean,
 and we presume that the strength of reversion might be different. This is because we expect
that  whether the  price strongly deviates from the mean value, as during a spike,
then it returns to  the average level with a stronger force than
usual.  

\medskip

Regarding the base component, let us note that in  many time
series of the electricity markets, an evidence of correlation
among price increments is clear. For example, see Figure
\ref{fig:autocorrelation_whole_and_calibration_window}. In order to capture better such a correlation within
different returns,  we consider an additive noise driven by a
fractional Brownian motion. 

\medskip

Furthermore,  the Italian market  is
rather peculiar since clearly identifiable spikes are rare; as a
consequence the intensity of the spike process is small and
becomes difficult to be estimated. Moreover, despite the small
number of spikes, a clustering effect seems to be present; so one might better include the effect of a self-exciting
stochastic process. Hence, by following recent literature
\cite{bacry_2014,bacry_2015,becker_clements_nur_2013,chen_2018,da_fonseca_2014,hainaut_2017,sgarra_2018},
we model the jump component $X_2$ via a Hawkes marked process.

\medskip

To be more precise, let us consider  a common filtered probability space
$(\Omega, \mathcal{F}, \{ \mathcal{F}_t\}_{t\in \mathbb R_+}, P)$. We suppose that $X_1$ follows a stochastic differential equation
driven by a fractional Brownian
motion $B^H=\{B^H(t), t\in \mathbb R_+\}$   with Hurst parameter
$H\in (0,1)$ and diffusion coefficient $\sigma \in \mathbb R_+$,
subject to mean reversion around a level zero, with strength
$\alpha_1 \in \mathbb R_+$.

\medskip
A Fractional Brownian motion (fBm) $(B^H_t)_{t\in \mathbb R_+}$
with Hurst parameter $H\in (0,1)$ is a   zero mean Gaussian
process with covariance function given by
$$
cov(B^H_s,B^H_t)=  \mathbb
E\left[B^H_tB^H_s\right]=\frac{1}{2}\left(|t|^{2H}+|s|^{2H}-|t-s|^{2H}\right).
$$

\medskip

The parameter $H$ is resonsible of the strenght and the sign of the correlations between the increments. Indeed, for $H\in(0,1)\setminus\{\frac12\}$,  set $\widetilde{H}=H-\frac{1}{2}$, for any $t_1<t_2<t_3<t_4$, one may express the covariance of the increment in an integral form
\begin{equation}
\label{eq: correlation of increments in integral form}
E\Big[(B^H_{t_2}-B^H_{t_1})(B^H_{t_4}-B^H_{t_3})\Big]=2 \widetilde{H} H\int_{t_1}^{t_2}\int_{t_3}^{t_4}(u-v)^{2\widetilde{H}-1}du\,dv ,
\end{equation}
Thus, since the integrand is a positive function, and $H>0$, the sign of the correlation depends only upon $\widetilde{H} $, being positive when $\widetilde{H}>0$, i.e. $H\in(\frac{1}{2},1)$, and negative when $\widetilde{H}<0$, i.e. $H\in(0,\frac{1}{2})$.

\medskip

For any $t\in \mathbb R_+$, we define $X_1(t)$  as the solution of the
following   equation
\begin{eqnarray}\label{eq_X1_initial}
dX_1(t)&=& - \alpha_{1}  X_1(t)d t +\sigma dB^H(t).
\end{eqnarray}

\begin{proposition}[\cite{kubilius book}]
	\label{prop:existence_X_1} Given $\alpha_1, \sigma \in \mathbb
	R_+$ and $H\in (0,1)$, Equation \eqref{eq_X1_initial}  admits
	the unique solution
	\begin{eqnarray}
	\label{eq:solution_fractional_ornstein-Uhlenbeck_process}
	X_1(t)&=&X_1(0) \,e^{-\alpha_1 t} -\alpha_1 \,\sigma e^{-\alpha_1
		t}\int_0^t e^{-\alpha_1 s} d B^H(s) +\sigma B^H(t)\\ &=&X_1(0)
	\,e^{-\alpha_1 t} +\sigma  \int_0^t e^{-\alpha_1 (t-s)} d
	B^H(s)).\nonumber
	\end{eqnarray}
	$X_1$ is called a fractional Ornstein-Uhlenbeck
	process.\end{proposition} The covariance structure of such a
process is rather complex (see Theorem 1.43 in
\cite{kubilius book}), simplified in the case of the variance of
the 1-dimensional marginals.
\begin{proposition}[\cite{kubilius book}, Theorem 1.43]
	\label{prop:properties_X_1} Given $\alpha_1, \sigma \in \mathbb
	R_+$ and $H\in (0,1)$, the following properties hold.
	
	\begin{itemize}
		\item[i)]For any $t\in \mathbb R_+$, the variable $X_1(t)$ has a
		Gaussian distribution, i.e.
		\begin{equation}
		\label{eq:X_1_distribution_normal} X_1(t) \sim
		\mathcal{N}\left(X_1(0) e^{-\alpha_1 t}, V_{\alpha_1}(t) \right),
		\end{equation}
		where the variance $V_{\alpha_1}(t)$ is given by
		\begin{equation}
		\label{eq:Variance_X_1_distribution_normal}
		V_{\alpha_1}(t)=H\sigma^2 \int_0^t x^{2H-1}\left(e^{-\alpha_1 s} +
		e^{-\alpha_1 (2t-s)} \right)  ds.
		\end{equation}
		
		\item[ii)] The variance has the following time asymptotic behavior
		
		$$
		\lim_{t\rightarrow +\infty} V_{\alpha_1}(t) = \alpha_1^{-2H}\,H
		\sigma^2 \,\Gamma(2H),
		$$
		where $\Gamma: \mathbb R \rightarrow \mathbb R$ is the classical
		$\Gamma$ function.
	\end{itemize}
\end{proposition}

\medskip

We move now to the $X_2$ component: we wish to define it as
the solution of a mean-reverting SDE driven by a Hawkes marked process $\pi$,  i.e.\ as 
\begin{equation}
\label{eq: jump component}	
X_2(t)=X_2(0)- \int_0^t  \alpha_{2}   X_2(s) ds  + \int_0^t
\int_0^{\lambda_s} \int_0^\infty z\,\,\pi\left(ds,d\eta,dz\right).
\end{equation}
We introduce its components in detail. Consider a marked point process
\begin{equation} \label{eq:(T,Z)}
\{ (T_i, Z_i)\}_{i\in \mathbb N},
\end{equation}
where, for any $ i \in \mathbb N$, $T_i$ is the random time at
which the $i$-th jump occurs and  $Z_i$ is the relative random
jump size. So we may express  the counting  measure $J$ of the jumps  via the marked process \eqref{eq:(T,Z)} as
\begin{equation} \label{eq:Jt}
J(dt)=\sum_{i=1}^\infty Z_i
\,\,\epsilon_{T_i}(dt)=\int_{\mathbb{R}} z Q(dt,dz),
\end{equation}
where $\epsilon_x$ is the Dirac measure localized in $x$ and $Q$ is
the following marked counting
measure on $\mathbb R_+\times \mathbb R$
\begin{equation} \label{eq:measure_Q}
Q(dt,dz)=\sum_{i=1}^\infty \,\,\epsilon_{(T_i,Z_i)}(dt,dz).
\end{equation}

\medskip

The counting process $ N=\{N_t\}_{t\in \mathbb R_+}$ associated
to the marked point process \eqref{eq:(T,Z)}
is such that, for any $t\in \mathbb R_+$
\begin{equation} \label{eq:Nt}
N_t=\sum_{i=1}^{\infty} \epsilon_{T_i}([0,t])=Q([0,t]\times
\mathbb R).
\end{equation}
The process $N$ is characterized by its time dependent conditional
intensity $\lambda_t$, $t\in \mathbb R_+$, which is the quantity such that:
$$
\lambda_t = \lim_{dt\rightarrow 0} \frac{\mathbb E\left[ N_{t+dt}-N_t| \mathcal{F}_t \right]}{dt},
$$ and
$$
prob\left( N_{t+dt}-N_t=k| \mathcal{F}_t \right)=\left\{%
\begin{array}{ll}
1- \lambda_t \, dt + o(dt), &  {k = 0;} \\
\lambda_t \, dt + o(dt), &  {k = 1;} \\
o(dt), &  {k >  1.} \\

\end{array}%
\right.
$$

In our case, we suppose that, for any $t\in \mathbb R_+$, $\lambda_t$ is a
function of past jumps of the process, i.e.
\begin{equation}
\label{eq:intensity_Hawkes} \lambda_t= \lambda+ \int_0^t \Phi(t-s)
dN_s,
\end{equation}
with background intensity  $ \lambda >0$   and excitation function
$\Phi : \mathbb R_+ \rightarrow   \mathbb R_+ $. Whenever
$\Phi(\cdot)\not= 0$, the resulting process is different from a homogeneous Poisson process, and if
\begin{equation}\label{eq:Phi_less_1_stationarity}
\|  \Phi\|_1=\int_0^\infty \Phi(t) dt < 1,
\end{equation}  the existence of a unique process is
implied. Condition \eqref{eq:Phi_less_1_stationarity} also implies the
stationarity of the process, that is that its distributions are
invariant under translations \cite{bremaud_Massoulie_1996,bacry_2014,bacry_2015}.
Equation
\eqref{eq:intensity_Hawkes} states that   the random
times of the jumps are governed by a constant intensity $\lambda$ and
that any time a jump occurs,   it excites the process in the sense
that   it changes the  rate of the arrival of subsequent jumps, by
means   of a kernel $\Phi$. Usually, the latter decreases to 0, so
that the influence of a jump upon future jumps decreases and tends
to 0 for larger time increments. We say in this case that $N$ is a univariate
Hawkes processes \cite{hawkes_1971_1,hawkes_1971_2}.
\\
Note that we may make explicit the dependence of the intensity
process upon the random jump times $\{T_i\}_{i\in \mathbb N}$ by
the following
$$
\lambda_t= \lambda+ \int_0^t \Phi(t-s) \sum_{i\in \mathbb N}
\epsilon_{T_i}(ds)=\lambda+ \sum_{i\in \mathbb N \,\,: \,\,\,
	T_i\le t}
\Phi(t-T_i) .
$$

As it happens in many examples in modelling (see \cite{becker_clements_nur_2013,bacry_2014,bacry_2015}), we consider an exponential model for the excitation
function, that is
\begin{equation}
\label{eq:exitation_Phi_exponential} \Phi(t)=\gamma e^{-\beta t},
\end{equation}
where $\gamma, \beta \in \mathbb R_+$ represent the instantaneous
increase after a jump and the speed of the reversion
to $\lambda$ of the excitation intensity. As a consequence, the
intensity \eqref{eq:intensity_Hawkes} becomes
\begin{equation}
\label{eq:intensity_exp_Hawkes} \lambda_t= \lambda+ \int_0^t \gamma e^{-\beta (t-s)}
dN_s= \lambda+ \gamma \sum_{i\in \mathbb N \,\,: \,\,\,
	T_i\le t} e^{-\beta (t-T_i)}.
\end{equation}
This process may be seen as a solution of the following stochastic
differential equation

\begin{equation}
\label{eq:sde_intensity_exp_Hawkes}d \lambda_t= \beta \left(
\lambda - \lambda_t\right)  +  \gamma dN_s.
\end{equation}

Notice that \eqref{eq:intensity_exp_Hawkes} is the solution of the
equation \eqref{eq:sde_intensity_exp_Hawkes} when the process
starts in $\lambda_0$ infinitely in the past and it is at its
stationary regime.  Otherwise, in order to model a process from
some time after it is started and  setting an initial condition
$\lambda_0 = \lambda^*$  the conditional intensity, solution of
\eqref{eq:sde_intensity_exp_Hawkes}   would be
\begin{equation}
\label{eq:intensity_exp_Hawkes_lambda_0}
\lambda_t= e^{-\beta t} \left(\lambda^*-\lambda\right)+\lambda+ \int_0^t \gamma e^{-\beta (t-s)}
dN_s.
\end{equation}

As mentioned above, for $t$ large enough the impact of the initial
condition vanishes, since the first term would die out. Note that a new jump of $N_t$
increases the intensity, which increases the
probability of new jump,  but the process does not necessarily blow up because
the drift is negative if $ \lambda_t > \lambda$. Furthermore, while the process $\{N_t
\}_{t\in \mathbb R_+}$ is non Markovian, the bidimensional process
$ \{(N_t,\lambda_t)\}_{t\in \mathbb R_+}$ is a Markov process
\cite{bacry_2015}, such that
\begin{eqnarray}\label{eq:hawkes_expected_N}
d\mathbb E\left[N_t\right] &=& \mathbb E\left[\lambda_t\right] dt,\\
d\mathbb E\left[\lambda_t\right] &=&  \left(\beta \lambda + (\gamma - \beta)\mathbb E\left[\lambda_t\right]\right) dt.
\label{eq:hawkes_expected_lambda}
\end{eqnarray}

Since the solution of equation \eqref{eq:hawkes_expected_lambda}
is
$$ \mathbb E\left[\lambda_t\right] = \mathbb
E\left[\lambda_0\right] e^{(\gamma-\beta)t}+ \beta\gamma\int_0^t
e^{-(\gamma-\beta)(t-s)}ds,
$$
if $\gamma>\beta$, then the intensity would explode in the
average, and so it would happen for the process $N_t$. This is not
the case in the stationary regime, since in the exponential case,
assumption \eqref{eq:Phi_less_1_stationarity} becomes $$ 1>\nu=\|
\Phi\|_1=\int_0^\infty \gamma e^{-\beta t}
dt=\frac{\gamma}{\beta},
$$ i.e.
\begin{equation}
\gamma < \beta. \label{eq:gamma_less_beta}
\end{equation}

\medskip
 
With this definition of $N$ in mind, we introduce the noise term $\pi$ appearing in the stochastic differential
equation \eqref{eq: jump component} that defines the process $X_2$. Let $\pi$ be a Poisson random
measure on $\mathbb R_+\times \mathbb R_+ \times \mathbb R $
with intensity measure  $\Lambda=\nu_+  \times \nu_+ \times \mu, $
where $\nu_+ $ is  a Lebesgue measure on $\mathbb R_+.  $ The
measure $\mu$ is the distribution of the size of the jumps that
satisfies condition \eqref{eq:property_mu}. We suppose that the
size distribution is given by a Borel measure $\mu$   on $\mathbb
R_+$, satisfying the condition
\begin{equation}\label{eq:property_mu} \int_{ 0}^\infty (\eta
\wedge \eta^2)\mu(d\eta) <\infty.
\end{equation}

If we suppose $\mu(d\eta)=\epsilon_{1}(d\eta)$, the jumps are of
size one. In \cite{delattre_fournier_hoffman_2016}, in a more
general setting in which they consider multidimensional non linear
Hawkes process, the author prove that the Hawkes process
\eqref{eq:Nt} with conditional intensity given by
\eqref{eq:intensity_exp_Hawkes} may be written as
\begin{equation}
\label{eq:Nt_poisson_measure} N_t=\int_0^t
\int_0^{\lambda_t} \int_0^\infty \pi\left(ds,d\eta,dz\right).
\end{equation}

\medskip

In conclusion, the process $X=X_1 +X_2$ is given by the solution of the
following system, for $ t\in \mathbb R_+$,

\begin{eqnarray}
X_1(t)&=&X_1(0)-\int_0^t  \alpha_{1}   X_1(s) ds  +\sigma \int_0^t
dB^H(s) ;
\label{eq:X_1_fractional diffusion}\\
X_2(t)&=&X_2(0)- \int_0^t  \alpha_{2}   X_2(s) ds  + \int_0^t
\int_0^{\lambda_s} \int_0^\infty z\,\,\pi\left(ds,d\eta,dz\right),\label{eq:X_2_jump}
\end{eqnarray}
coupled with equation \eqref{eq:sde_intensity_exp_Hawkes}, with $
\gamma, \beta, \lambda \in \mathbb R_+$.

System \eqref{eq:X_1_fractional diffusion}-\eqref{eq:X_2_jump}
admits a unique solution thanks to the mentioned results.

\subsection{Path simulations}

\label{subsec: path simulation}

In the following we consider the simulation results showing the
macroscopic behaviour of the model, by considering some fixed set of parameters in
Tables
\ref{tab:simulation_fixed_parameters}--\ref{tab:simulation_H_beta_gamma}.
For the jump size distribution $\mu$ we choose to consider the Generalized Extreme value
distribution, that is a probability measure depending from 3 parameters $\tilde{\mu},\xi \in \R$ and $\tilde{\sigma} >0$ with density function
given by
\begin{equation}
\label{eq:laplace_distribution} 
f(x)= \frac{1}{\tilde\sigma}\,t(x)^{\xi+1}e^{-t(x)},
\end{equation}
where 
$${\displaystyle t(x)={\begin{cases}{\big (}1+\xi ({\tfrac {x-\tilde\mu }{\tilde\sigma }}){\big )}^{-1/\xi }&{\textrm {if}}\ \xi \neq 0\\e^{-(x-\tilde\mu )/\tilde\sigma }&{\textrm {if}}\ \xi =0.\end{cases}}}$$
We considered the following fixed set of parameters

\begin{table}[h]
	\centering
	{ \small
		\begin{tabular}{c|c|c|c|c|c|c}
			\hline
			$\alpha_1$  &   $\alpha_2$    & $\sigma$    & $\lambda $  & $\tilde\mu$ & $\xi $ &  $\tilde{\sigma}$\\
			\hline
			0.1 &  0.5 &  6 &    0.01 & 18 & 0.7 &  2  \\
			\hline
		\end{tabular}
	}
	\caption{\small{ Fixed parameters used in the simulations: mean-reverting parameters $\alpha_1$ and $\alpha_2$,
			diffusion coefficient $\sigma$, basic Poisson point process parameter $\lambda$ and the parameters $\tilde{\mu},\xi \in \R$ and $\tilde{\sigma} >0$
			in the Generalized Extreme Value distribution \eqref{eq:laplace_distribution}.}}
	\label{tab:simulation_fixed_parameters}
\end{table}
\medskip
On the other hand, we consider a changing set of parameters to evaluate the impact of some of the important features that we introduce with our model: the fBm depending from the parameter $H$ and the parameters $\gamma,\beta$ of the self-exciting part of the Hawkes process 
\begin{table}[h]
	\centering
	{ \small
		\begin{tabular}{c|c|c|c}
			\hline
			\textbf{Parameter  } &  \textbf{ a.1.  } & \textbf{ a.2.  }   &\textbf{ a.3.  }    \\
			\hline
			$H$  &  0.2 &     0.5 &  0.7 \\
			\hline
		\end{tabular}
	}
	\caption{\small{Set of simulation parameters: values for the  Hurst  parameter $H$
			in the diffusion term  in \eqref{eq:X_1_fractional diffusion}.}}
	\label{tab:simulation_H}
\end{table}
\medskip
\begin{table}[h!]
	\centering
	{ \small
		\begin{tabular}{c|c|c|c|c}
			\hline
			\textbf{Parameter  } &  \textbf{ (a)  } & \textbf{ (b) }   &\textbf{ (c)  }   & \textbf{ (d)  }  \\
			\hline
			$\gamma$  &  0 &  0.05 &    0.15 &  0.3 \\
			$\beta$  &  0 &  0.08 &    0.2 &  0.5 \\
			\hline
		\end{tabular}
	}
	\caption{\small{ Set of simulation parameters for the Hawkes excitation function:
			$\gamma$ and $\beta$ in \eqref{eq:sde_intensity_exp_Hawkes} satisfying stationarity condition \eqref{eq:gamma_less_beta}. }}
	\label{tab:simulation_H_beta_gamma}
\end{table}

\medskip

For this set of simulation we consider a fixed deterministic function $$ f(t)=130 \cdot
1_{[0,\infty)}(t),
$$
and the following deterministic initial condition for the processes $X_1$ and $X_2$
\begin{eqnarray*}
	X_1(0)&=& X_2(0) =0.
\end{eqnarray*}
Stochastic simulation are carried out by generating  exact paths of Fractional Gaussian Noise by using
circulant embedding (for $1/2 < H<1$) and Lowen's method (for
$0 < H < 1/2$), while the
Hawkes process is generated by a thinning procedure for
inhomogeneous Poisson process as in \cite{ogata}. 

\medskip

\begin{figure}[h!]
	\centering
	\includegraphics[width=\textwidth]{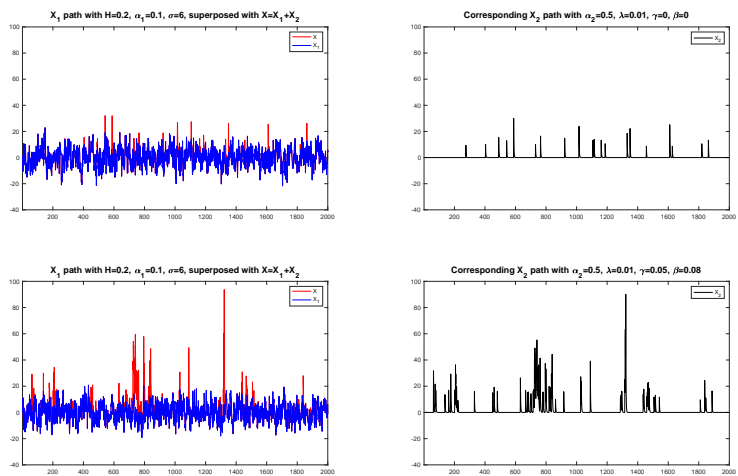}
	\caption{Path of the process $X=X_1+X_2$ (red) and of $X_1$ alone (blue), with the corresponding $X_2$ in black on the right. 
		Case $H=0.2$ and (a) $\lambda=0, \beta=0$; (b) $\lambda=0.05, \beta=0.08$}
	\label{fig:traiettoria fig1}
\end{figure}

\begin{figure}[h!]
	\centering
	\includegraphics[width=\textwidth]{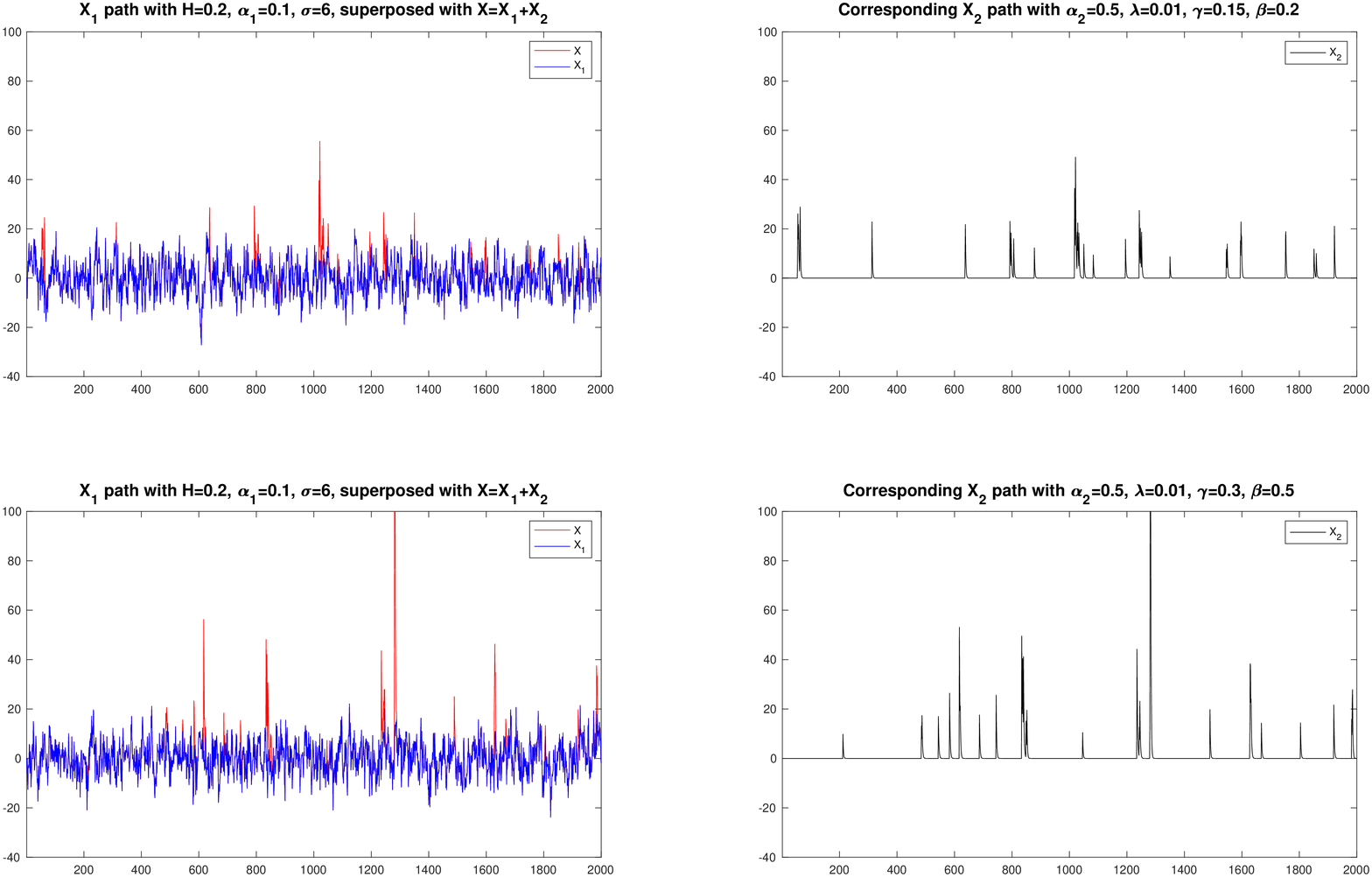}
	\caption{Path of the process $X=X_1+X_2$ (red) and of $X_1$ alone (blue), with the corresponding $X_2$ in black on the right. 
		Case $H=0.2$ and (c) $\lambda=0.15, \beta=0.2$; (d) $\lambda=0.3, \beta=0.5$.}
	\label{fig:traiettoria fig2}
\end{figure}

\begin{figure}[h!]
	\centering
	\includegraphics[width=\textwidth]{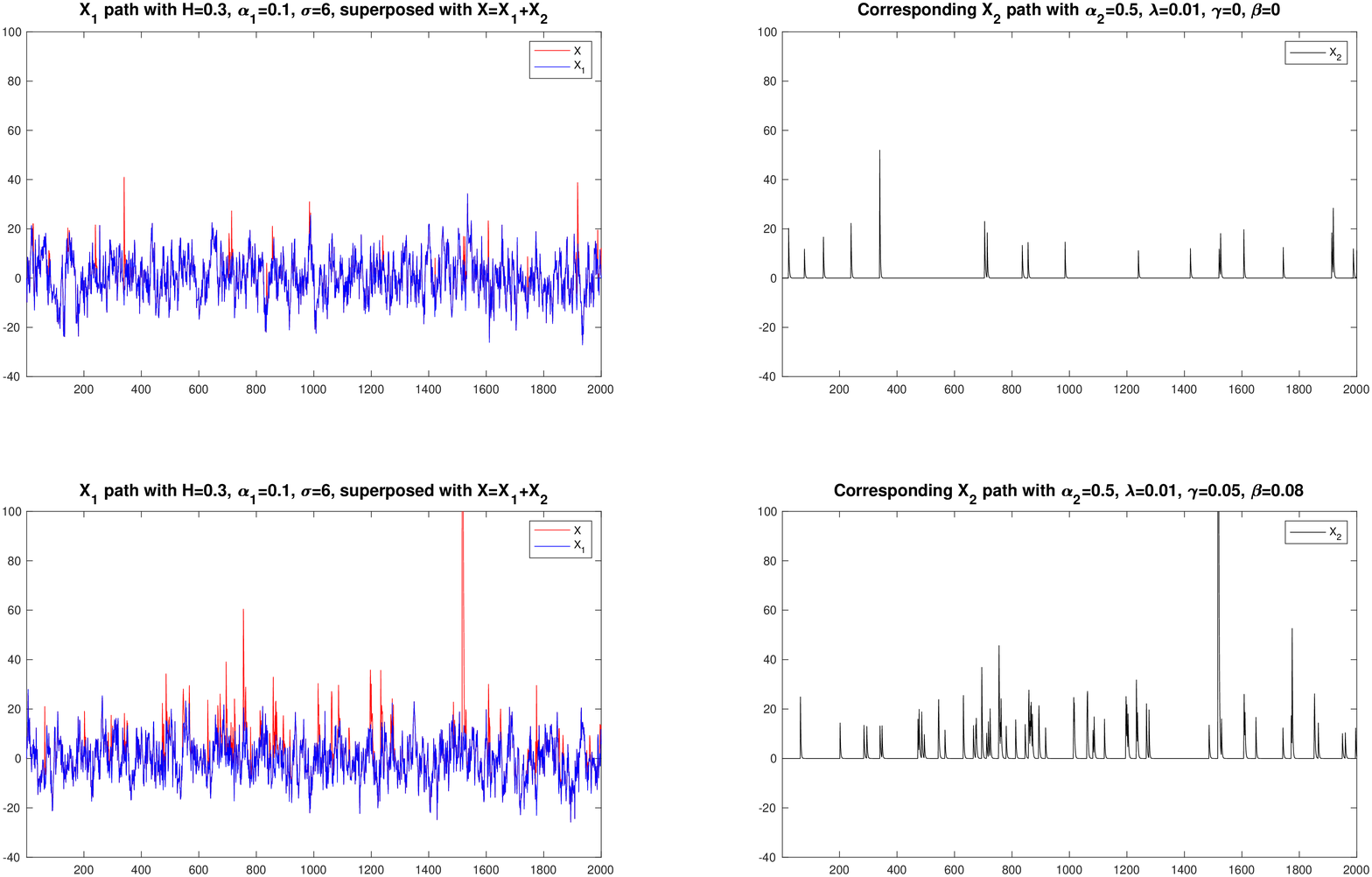}
	\caption{Path of the process $X=X_1+X_2$ (red) and of $X_1$ alone (blue), with the corresponding $X_2$ in black on the right. 
		Case $H=0.3$ and (a) $\lambda=0, \beta=0$; (b) $\lambda=0.05, \beta=0.08$}
	\label{fig:traiettoria fig3}
\end{figure}

\begin{figure}[h!]
	\centering
	\includegraphics[width=\textwidth]{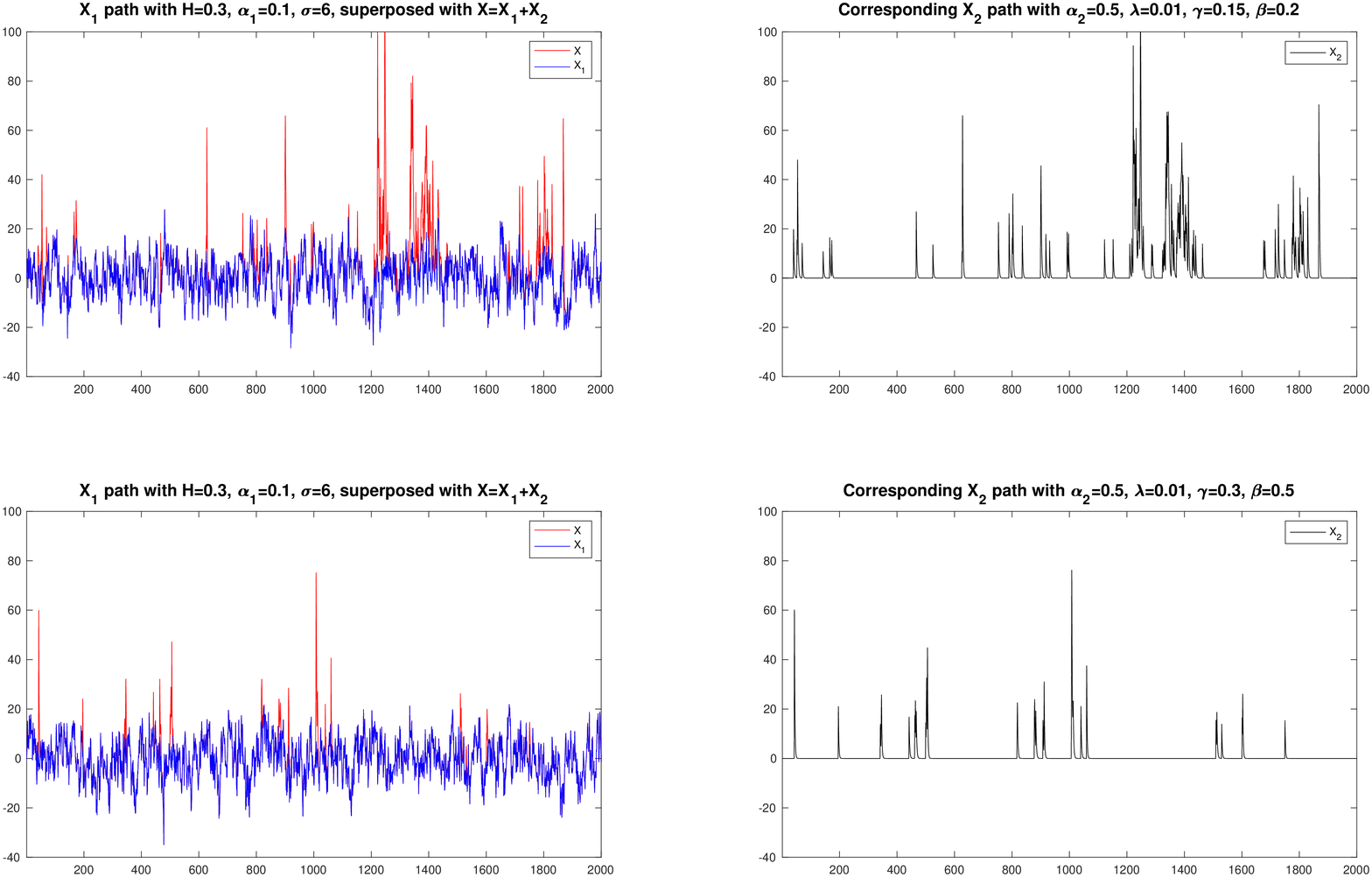}
	\caption{Path of the process $X=X_1+X_2$ (red) and of $X_1$ alone (blue), with the corresponding $X_2$ in black on the right. 
		Case $H=0.3$ and (c) $\lambda=0.15, \beta=0.2$; (d) $\lambda=0.3, \beta=0.5$.}
	\label{fig:traiettoria fig4}
\end{figure}

\begin{figure}[h!]
	\centering
	\includegraphics[width=\textwidth]{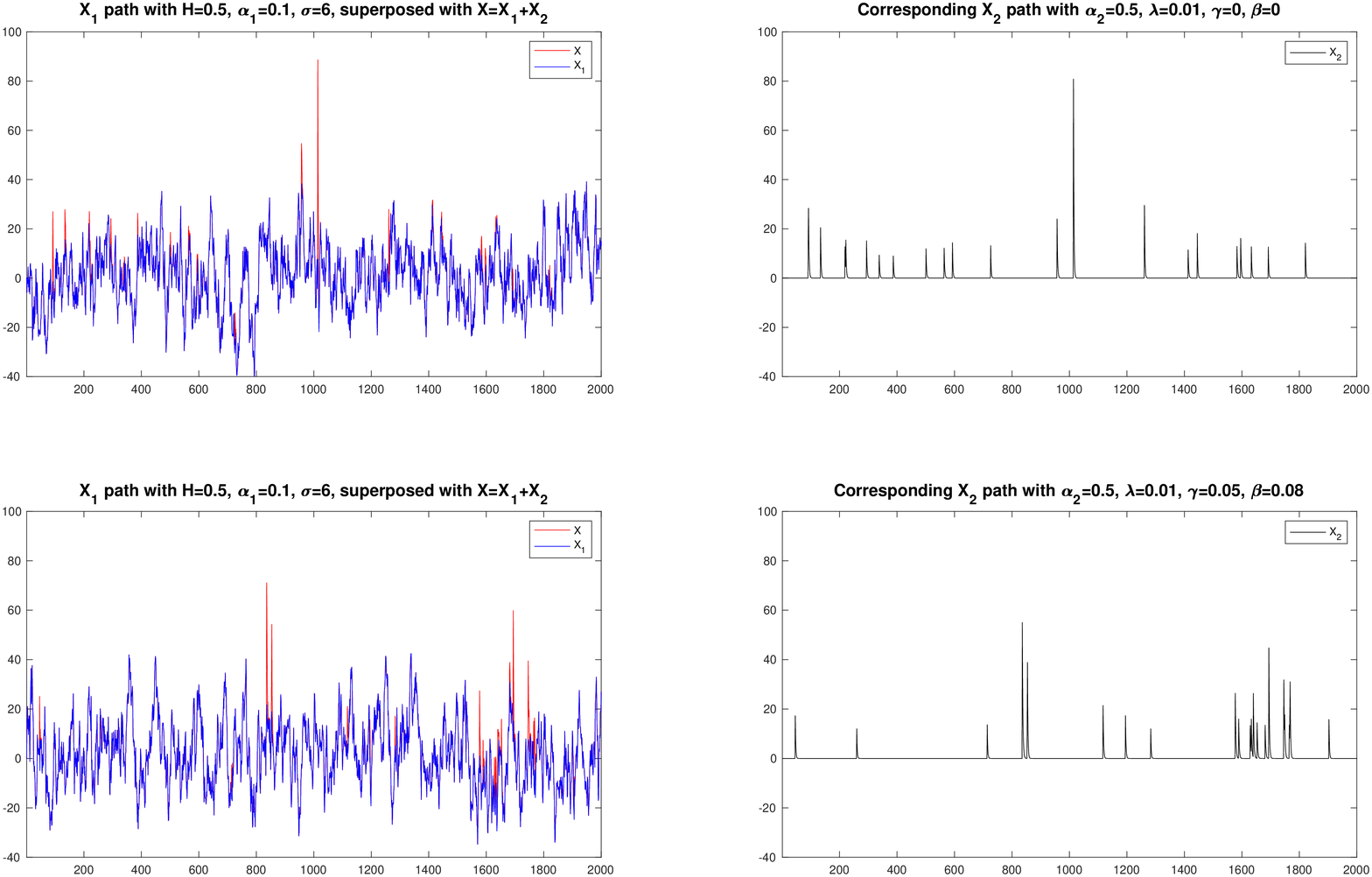}
	\caption{Path of the process $X=X_1+X_2$ (red) and of $X_1$ alone (blue), with the corresponding $X_2$ in black on the right. 
		Case $H=0.5$ and (a) $\lambda=0, \beta=0$; (b) $\lambda=0.05, \beta=0.08$}
	\label{fig:traiettoria fig5}
\end{figure}

\begin{figure}[h!]
	\centering
	\includegraphics[width=\textwidth]{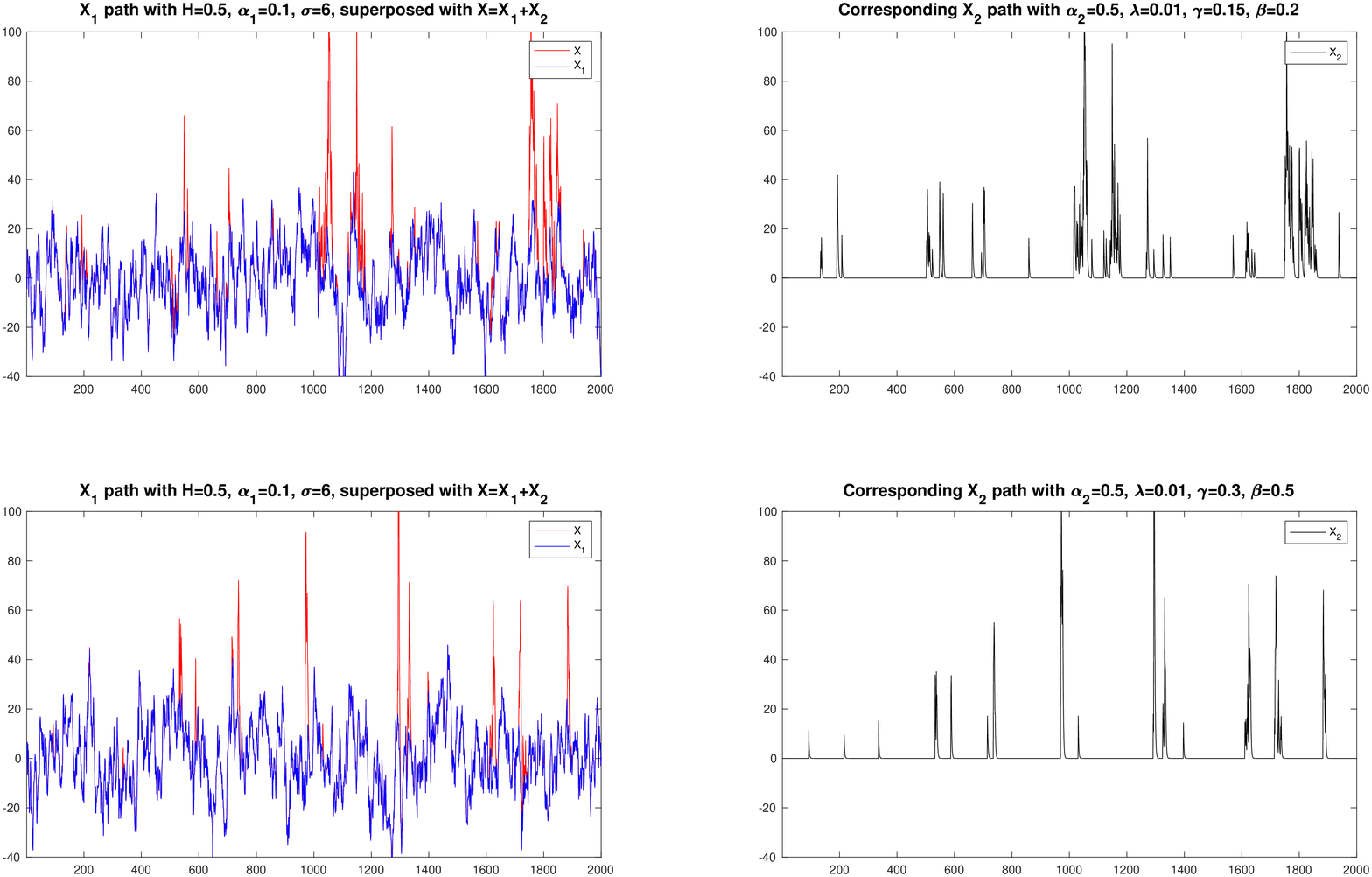}
	\caption{Path of the process $X=X_1+X_2$ (red) and of $X_1$ alone (blue), with the corresponding $X_2$ in black on the right. 
		Case $H=0.5$ and (c) $\lambda=0.15, \beta=0.2$; (d) $\lambda=0.3, \beta=0.5$.}
	\label{fig:traiettoria fig6}
\end{figure}

\begin{figure}[h!]
	\centering
	\includegraphics[width=\textwidth]{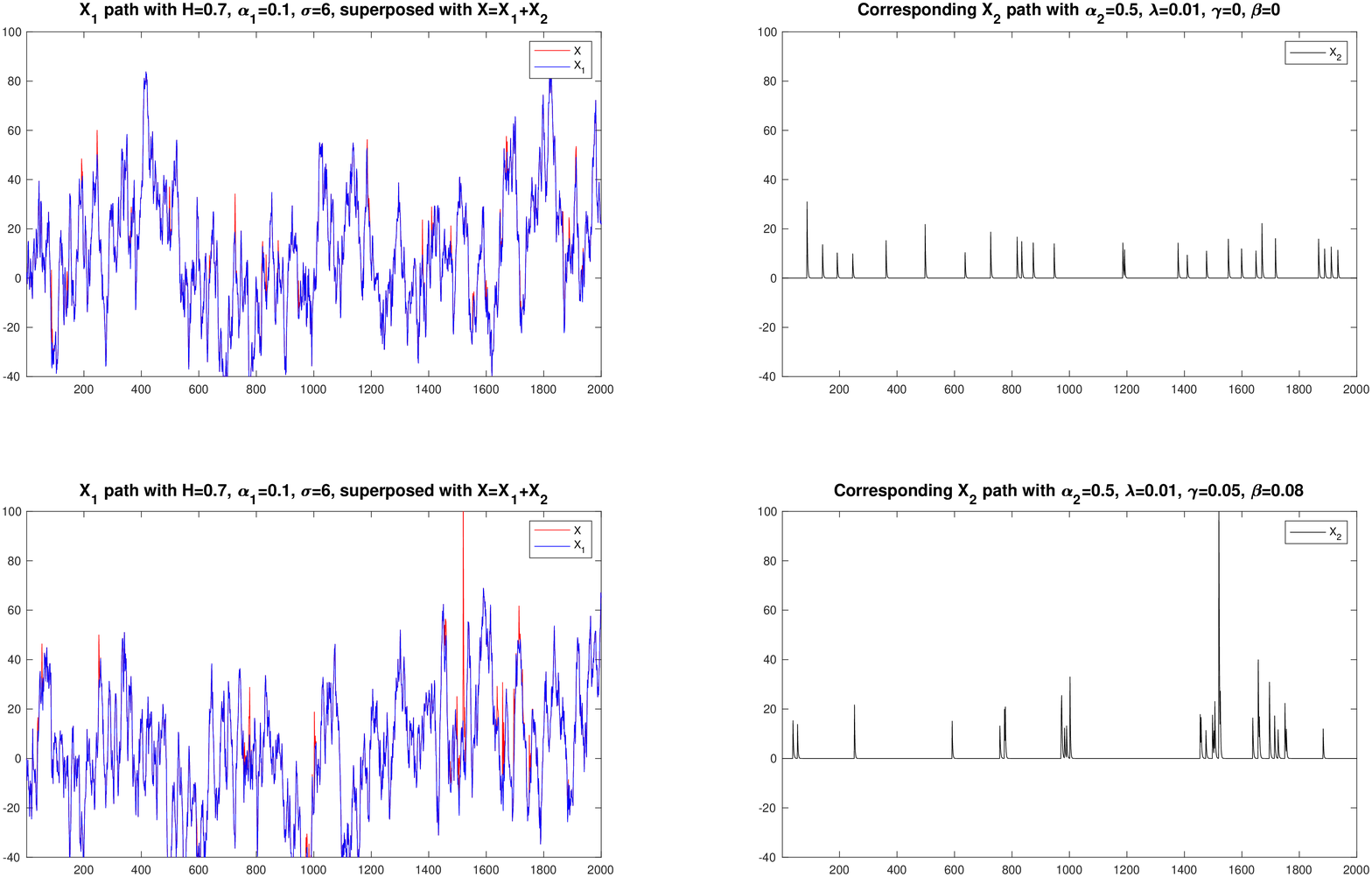}
	\caption{Path of the process $X=X_1+X_2$ (red) and of $X_1$ alone (blue), with the corresponding $X_2$ in black on the right. 
		Case $H=0.7$ and (a) $\lambda=0, \beta=0$; (b) $\lambda=0.05, \beta=0.08$}
	\label{fig:traiettoria fig7}
\end{figure}

\begin{figure}[h!]
	\centering
	\includegraphics[width=\textwidth]{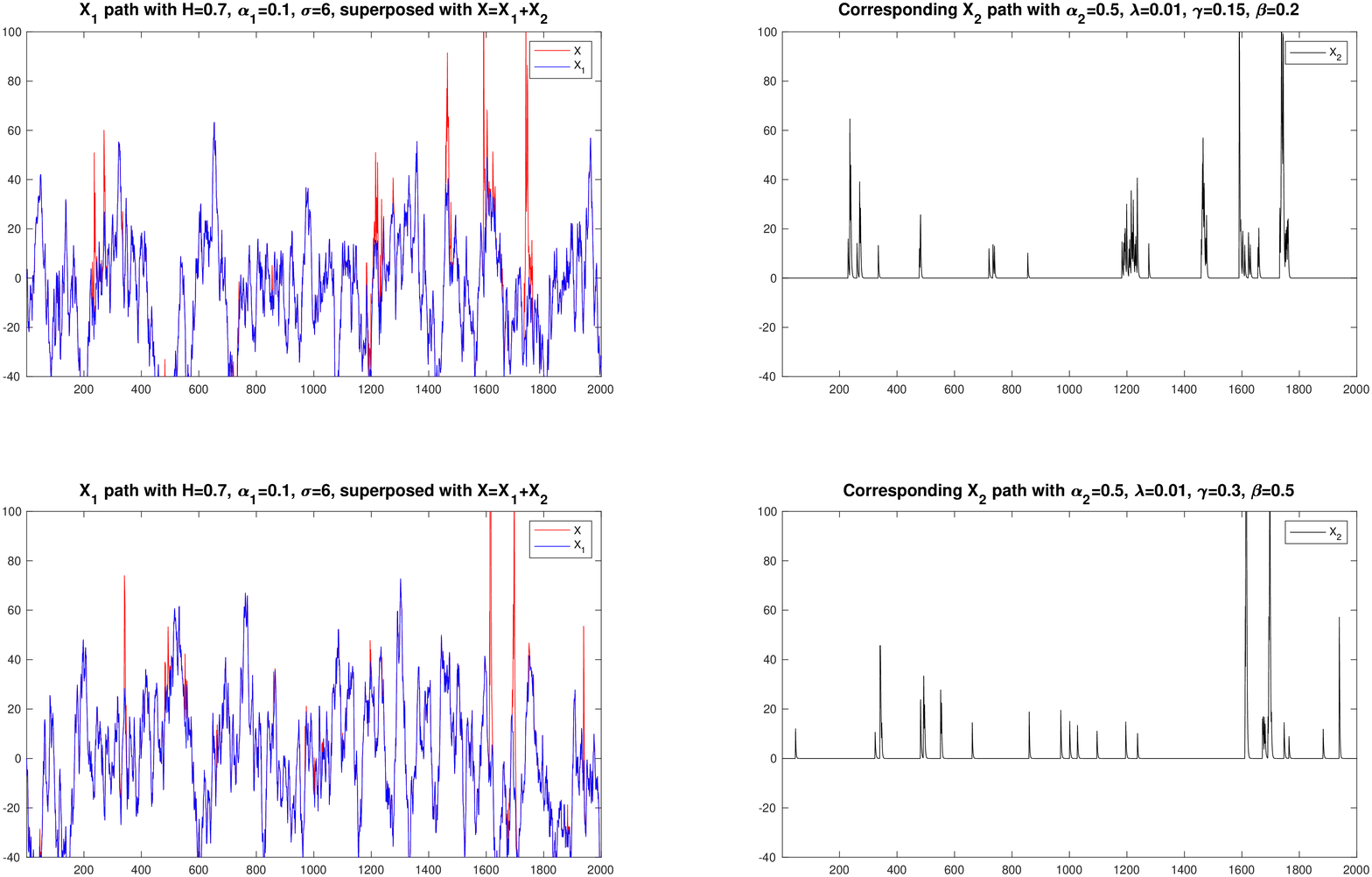}
	\caption{Path of the process $X=X_1+X_2$ (red) and of $X_1$ alone (blue), with the corresponding $X_2$ in black on the right. 
		Case $H=0.7$ and (c) $\lambda=0.15, \beta=0.2$; (d) $\lambda=0.3, \beta=0.5$.}
	\label{fig:traiettoria fig8}
\end{figure}

 Figures \ref{fig:traiettoria fig1}--\ref{fig:traiettoria fig8} show some simulations of a single path of $X=X_1+X_2$ for the different values of the parameters chosen in Tables \ref{tab:simulation_fixed_parameters}--\ref{tab:simulation_H_beta_gamma}.
 
 Even if in some cases this meant that some parts of the path are not visible, we chose to keep the same scale in all figures. This makes us see very clearly the differences caused in the nature of the process $X_1$ by the changes in the values of $H$. We see in particular that, keeping $\alpha_{1},\sigma$ fixed we get a much more variable process as long as $H$ increases. 

\medskip

Regarding the jump component $X_2$, which is independent from $X_1$, we see that the cluster effect is clearly visible for the sets of parameters (b)--(d). It seems that the set of parameters (b) is producing more clusters than the others. This may seem strange, since in this case the parameter $\gamma$ is lower than in (c) and (d), but we remark that in all cases the parameter $\beta$, which models the speed of mean reversion of $\lambda_t$ towards $\lambda$, varies consistently with $\gamma$.

\medskip

We make a remark about the relation of this simulations with the real data. If we compare Figures \ref{fig:traiettoria fig1}--\ref{fig:traiettoria fig8} with Figure \ref{fig: rough data plot}, in which we plot the entire dataset that we will analyse, we can make some qualitative considerations. Our simulations of $X_1+X_2$ do not include any seasonal component, and this is clearly visible. Anyway, from the point of view of the appearance of the paths, we see some similarities between Figure \ref{fig: rough data plot}, the bottom plot of Figure \ref{fig:traiettoria fig3} and the top plot of Figure \ref{fig:traiettoria fig4}, which are relative to $H=0.3$ and the set of parameters (b) and (c) for the Hawkes process, corresponding to $\gamma=0.05, \, \beta=0.08$ and $\gamma=0.15,\, \beta=0.2$. In both of these figures the standard volatility seems quite similar to the one of the real data, and moreover the jump behaviour (both amplitude and clusters) is quite similar to the one of Figure \ref{fig: rough data plot}. 

\smallskip

These considerations are coherent with the estimates that we will get in Section \ref{sec: forecasting performance}.

\section{Parameter estimation}

\label{subsec: parameter estimation}

This section is devoted to the methods of estimation for the
parameters of the two stochastic components $X_1$ and $X_2$ of our
model. We recall that we have the following set of parameters to
be estimated from the dataset $Y$

\begin{table}[h]
	\centering
	{ \small
		\begin{tabular}{|c|c|}
			\hline
			\textbf{Equation} & \textbf{Parameter} \\
			\hline
			\hline
			$\mathbf{X_1}$   &  $\alpha_1$ \\
			\hline
			$\mathbf{X_1}$  & $\sigma$  \\
			\hline
			$\mathbf{X_1}$  & $H$  \\
			\hline
			\hline
			$\mathbf{X_2}$   & $\alpha_2$  \\
			\hline
			$\mathbf{X_2}$   & $\lambda_0$  \\
			\hline
			$\mathbf{X_2}$ & $\gamma$ \\
			\hline
			$\mathbf{X_2}$  & $\beta$  \\
			\hline
			$\mathbf{X_2}$  & parameters for jump distribution \\
			\hline
		\end{tabular}
	}
	\caption{Parameters of the model}
\end{table}

\subsection{Estimations for $X_1$}

For the base component $X_1$, we have to estimate three
parameters: $\alpha_1$, which is the rate of mean-reversion of the
process to the zero level, $\sigma$, which is the variance
parameter of the fractional noise $B^H$, and the parameter of
fractionality of the noise $H$ itself, which determines whether
the contributions of the noise term are positively correlated (if
$H>1/2$), negatively correlated (if $H<1/2$) or non-correlated (if
$H=1/2$, which is the case in which the noise is a classical
Brownian motion).

\medskip

We discuss first the estimation of the Hurst exponent $H$.
Its estimation  is very important   since it determines the magnitude of the
self-correlation of the noise of our model.  There are various
techniques that can be used to infer the Hurst coefficient from a
discrete signal. In \cite{Hurst revisited}  the reader may find  a good reviewon the argument. Anyway, these techniques can be used to estimate the Hurst
coefficient only by assuming that the observed path belongs only to  a
fractional Brownian motion $\{B_t^H(\omega)\}_t$, without
any drift. So it is not our case.
We use the estimator provided in
\cite{kubilius book}, given the observations of a wide class of SDEs driven by fractional Brownian motion, including the case of an Ornstein-Uhlenbeck model.

Given the time interval of observation   $[0,T]$, let us consider a partition
 $\pi_n=\left\{ T\displaystyle \frac{k}{n}\right\}_k $, where ${k\in\{0,1,\ldots,n\}}$.
Given a stochastic process $\{ X_1\}_{t\in[0,T]}$, let us define the two step increment as follows
$$\Delta^{(2)}_{n,k}X_1:=X_1\left(\frac{k+1}{n}T\right)-2X_1\left(\frac{k}{n}T\right)
+X_1\left(\frac{k-1}{n}T\right).$$
By Theorem 3.4 of \cite{kubilius book}, we have that the statistics
$$\hat{H}_n:=\frac{1}{2}-\frac{1}{2 \log 2} \log \Bigg(  \frac{\sum_{k=1}^{2n-1} (\Delta^{(2)}_{2n,k}X_1)^2 }{\sum_{k=1}^{n-1} (\Delta^{(2)}_{n,k}X_1 )^2 } \Bigg)$$
is a strongly consistent estimator  of $H$, i.e.\ it
holds
$$
\lim\limits_{T \to\infty} \hat{H}_n=H,   a.s. 
$$
In our case, our dataset will be discretized in time steps with
minimum length equal to one, so we will have to stop at a specific
step of the discretization, which is exactly the maximum value of
$n$ such that $\frac{T}{2n}>1$. 
\medskip

The estimation of the diffusion cpefficient  $\sigma$ is based on the use of the $p$-the variation. Indeed  for any $k\geq 2$ and for any
$p\geq 1$, the estimator is defined as
\begin{equation}
\label{eq: estimator sigma}
\hat{\sigma}_T(n):=\frac{n^{-1+pH}V^n_{k,p}X_1(T)}{c_{k,p}T},
\end{equation}
where $V_{k,p}^nX_1(t)$ is the $k$-th order $p$-th variation
process defined by
\begin{eqnarray*}
V_{k,p}^nX_1(t)&:=&\sum_{i=1}^{[nt]-k+1} \Big|\Delta_k X_1\Big(\frac{i-1}{n}\Big)\Big|^p\\
&:=& \sum_{i=1}^{[nt]-k+1} \Bigg|  \sum_{j=0}^{k}  (-1)^{k-j} \binom{k}{j} X_1\Big(\frac{i+j-1}{n}\Big)  \Bigg|^p.
\end{eqnarray*}
The constant $c_{k,p}$   in \eqref{eq: estimator sigma}  is given by
$$c_{k,p}:=\frac{2^{p/2}\Gamma\Big( \frac{p+1}{2} \Big)}{\Gamma\Big(\frac{1}{2}\Big)}\rho_{k,H}^{p/2},$$
where $\rho_{k,H}$ is
$$\rho_{k,H}:=\sum_{j=-k}^{k} (-1)^{1-j}\binom{2k}{k-j}|j|^{2H}.$$
The estimator \eqref{eq: estimator sigma}   has been introduced in
\cite{nualart}. 
With a discrete dataset of time lenght $N$ and with minimum
discretization length equal to $1$, we can only compute the
estimator $\hat{\sigma}_T(n)$ for $n=1$ and $T=N$. We choose to
use $k=2$ and $p=2$.

\medskip

Finally, the mean reverting parameter $\alpha_1$ has been estimated with the following ergodic estimator 
\begin{equation}
\label{eq: estimator alpha continuous}
\hat{\alpha}_1(T):=\Big(  \frac{1}{\sigma^2 H \Gamma(2H)T} \int_{ 0}^{T} X_1(t)^2dt   \Big)^{-\frac{1}{2H}}.
\end{equation}
It holds  that \cite{nualart} $$
\lim\limits_{T\to \infty} \hat{\alpha}_1(T)= \alpha_1,  \quad a.s.$$  The estimator \eqref{eq: estimator alpha continuous} is continuous version, but it   can be easily discretized. Suppose
that we are observing our process $X_1$ in the time points
$\{kh\}$, for $h=0,\dots,n$. Here $h$ is the amplitude of the time
discretization, and we suppose that $h=h(n)$ is such that $hn\to
\infty$ and $h(n)\to 0$ when $n\to\infty$. In \cite{nualart} the
authors define the discretized estimator as
\begin{equation}
\label{eq: estimator alpha discrete}
\hat{\alpha}_1(n):=\Big(  \frac{1}{\sigma^2 H \Gamma(2H)T} \sum_{k=1}^{n} X_1(kh)^2   \Big)^{-\frac{1}{2H}}.
\end{equation}
The only difference with the continuous version is the
discretization of the integral appearing in \eqref{eq: estimator alpha continuous}. The following
result holds:
\begin{theorem}[\cite{nualart}, Theorem 5.6]
	\label{th: nualart}
	Let $X_1$ be the solution of an Ornstein-Uhlenbeck process observed at times $\{kh, k=0,\dots, n\}$, and such that $h=h(n)$ satisfies $hn\to \infty $ and $h(n)\to 0$ when $n\to \infty $. Moreover, if we suppose that
	\begin{itemize}
		\item[i)] if $H\in(0,\frac{3}{4})$, $nh^p\to 0$ as
		$n\to\infty$ for some $p\in
		\Big(1,\frac{3+2H}{1+2H}\wedge (1+2H)\Big)$
		\item[ii)] if $H=\frac{3}{4}$, $\frac{h^pn}{\log
			(hn)}\to 0$ as $n\to\infty$ for some
		$p\in(1,\frac{9}{5})$
		\item[iii)] if $H\in(\frac{3}{4},1)$, $h^pn\to 0$ as
		$n\to\infty$ for some $p\in\Big(
		1,\frac{3-H}{2-H} \Big)$
	\end{itemize}
	Then the estimator $\hat{\alpha}_1\xrightarrow{n\to\infty} \alpha_1$ a.s.\
\end{theorem}

We observe that in our case we are not able to extend the total
length of the observation interval, since only a single time series is given. In addition, our dataset consists
in daily observations, so that the minimum time increment $h$ that
we can consider is $h=1$. We aim to define a proper function
$h=h(n)$ such that it satisfies the conditions of Theorem \ref{th:
	nualart} for every value of $H\in (0,1)$ (which is in principle
unknown).

Let $N$ be the length of our dataset. As a fundamental condition,
we want that $h(N)=1$, so that we are able to compute the $N$-th
approximation of our estimator with our data. We look for our
candidate $h$ within the class of functions
$$h(n)=\Big(\frac{N}{n}\Big)^\delta,$$
for some positive $\delta$ to be determined. All $h$ in this class
satisfy that $h(N)=1$. We need also need to have that $nh\to\infty$, as
$n\to\infty$, which imposes the condition $\delta <1$. Moreover,
we have to impose on $h$ conditions \textit{i), ii), iii)} of
Theorem \ref{th: nualart}.  We compute
$$nh(n)^p=\frac{N^{p\delta}}{n^{p\delta-1}}.$$
In condition \textit{i)} and \textit{iii)} we need $nh^p\to 0$ as
$n\to\infty$, while in \textit{ii)} we need $\frac{nh^p}{\log
	(nh)}\to 0$. Since $\log(nh)\to \infty$ by hypothesis, the
condition $nh^p\to 0$ is more restrictive and we impose it also in
\textit{ii)}. Since we do not need a priori which is the value of
$H$ of our fractional Ornstein-Uhlenbeck process $X_1$, we find a
$p=p(H)$ that is a good choice for any value of $H\in (0,1)$. One
can easily verify that $p(H)=1+H$ lies in all the admissible
intervals for $p$ in \textit{i), ii), iii)}. With this choice of
$p$, the expression of $nh^p$ reads
$$nh(n)^p=\frac{N^{(1+H)\delta}}{n^{\delta(1+H)-1}}.$$
In order for the right-hand side to converge to zero we must have
$\delta>\frac{1}{(1+H)}$. So we are left with the pair of
conditions
$$\frac{1}{1+H}<\delta <1,$$
which are both satisfied if we define
$$\delta=\delta(H):=\frac{1}{(1+H)^{\frac{1}{2}}},$$
for any $H\in (0,1)$. With this choice of $h$, we are able to
calculate the $N$-th step of the approximation of $\alpha_1$,
regardless of the estimation $\hat{H}$ of $H$ that we obtained.
Furthermore, in the definition of $\hat{\alpha}_1(n)$ there is a clear
dependence on $\sigma$, which in our case is unknown. We remark
that anyway, since the estimator $\hat{\sigma}$ converges a.s.\ to
$\sigma$ as the order of the approximation
increases, we have that the estimator $\hat{\alpha}_1$ converges
a.s.\ to $\alpha_1$ also if we substitute $\hat{\sigma}$ to
$\sigma$ in its definition.

\begin{table}[h]
	\centering
	{ \small
		\begin{tabular}{|c||c|c|c| }
			\hline
			\textbf{Case} & $\hat{\alpha}_1$ & $q_{5\%}(\hat{\alpha}_1)$ & $q_{95\%}(\hat{\alpha}_1)$  \\
			\hline
			\hline
			a.1 & \textbf{0.1069} & 0.0050 & 0.2523  \\
			\hline
			a.2  & \textbf{0.1030}  & 0.0220  & 0.2098  \\
			\hline
			a.3  & \textbf{0.1017}  & 0.0445 & 0.1737\\
			\hline
			a.4   & \textbf{0.1019}  & 0.0531 & 0.1641 \\
			\hline
			\hline
			\textbf{Case}  & $\hat{\sigma}$ &$q_{5\%}(\hat{\sigma})$ & $q_{95\%}(\hat{\sigma})$    \\
			\hline
			\hline
			a.1 &    \textbf{6.2334} & 5.7834 & 6.7926   \\
			
			a.2  &     \textbf{6.2313} & 5.7705  & 6.8007 \\
			
			a.3  &       \textbf{6.2233} & 5.7024 & 6.8636 \\
			
			a.4  &     \textbf{6.1976}  & 5.4920 & 7.0783  \\
			\hline
			\textbf{Case}    & $\hat{H}$ &  $q_{5\%}(\hat{H})$ & $q_{95\%}(\hat{H})$  \\
			\hline
			\hline
			a.1 &    \textbf{0.1940} & 0.0885 & 0.2969     \\
			
			a.2  &     \textbf{0.2927} & 0.1902 & 0.3931  \\
			
			a.3  &     \textbf{0.4909} & 0.3982 & 0.5819  \\
			
			a.4  &      \textbf{0.6882}& 0.6024 & 0.7715 \\
			\hline
		\end{tabular}
	}
	\caption{Estimated parameters of the component $X_1$, given $M=20000$ realizations of the component $X_1$ itself for each of the parameters set a.1 -- a.4.}
\end{table}

\subsection{Estimations for $X_2$}

For the jump component $X_2$, there are three separated tasks to
carry out in order  to estimate the parameters of the model. First, one has
to estimate the parameters of the self-exciting intensity of the
Hawkes process. Second, one has to choose and fit an adequate
distribution for the jump magnitude. Third, one has to estimate
the mean-reverting parameter $\alpha_2$ appearing in
\eqref{eq:X_2_jump}.

We start with the parameters of the jump intensity $\lambda_t$
defined in \eqref{eq:intensity_exp_Hawkes}. In \cite{Ozaki} the
author gives an explicit formula for the log-likelihood of
$\lambda,\gamma,\beta$, the parameters of the intensity function $\lambda_t$, given  the
observed jump times $T_i$. The
log-likelihood takes the form
\begin{eqnarray*}
	L(T_1,\dots T_n| \lambda,\gamma,\beta)&=& -\lambda T_n+\sum_{j=1}^{n}\frac{\gamma}{\beta}\left(e^{-\beta(T_n-T_j)}-1\right)\\
	&&+\sum_{j=1}^{n} \log\Big(\lambda+\gamma A(j)\Big),
	\end{eqnarray*}
where $A(1)=0, A(j)=\sum_{i=1}^{j-1} e^{-\beta(T_j-T_i)}, j\ge 2$.
In order to have a more efficient maximization process, one can
immediately compute the partial derivatives of $L$. One has
\begin{eqnarray*}
	\frac{\partial \log L}{\partial \gamma}	 &=&\sum_{j=1}^{n}
	\frac{1}{\beta}\left(e^{-\beta(T_n-T_j)} -1\right)+\sum_{i=j}^{n}
	\left[\frac{A(i)}{\lambda+\gamma A(i)}\right];
	\\
\frac{\partial \log L}{\partial \beta}	 &=&- \gamma \sum_{j=1}^{n} \left[ \frac{1}{\beta} (T_n-T_j)e^{-\beta(T_n-T_j)}+ \frac{1}{\beta^2}  e^{-\beta(T_n-T_j)} \right]\\
&&-\sum_{j=1}^{n} \left[\frac{\gamma B(i)}{\lambda +\gamma A(i)} \right];
\\
\frac{\partial \log L}{\partial \lambda}	 &=&-T_n +  \sum_{j=1}^{n}\left[\frac{1}{\lambda+\gamma A(i)}\right];
\end{eqnarray*}
with
\begin{eqnarray*}
	\frac{\partial^2 \log L}{\partial \gamma^2}	 &=&-\sum_{i=j}^{n}
	\left[\frac{A(i)}{\lambda+\gamma A(i)}\right]^2;
	\\
		\frac{\partial^2 \log L}{\partial \beta^2}	 &=&\gamma
	 \sum_{j=1}^{n} \left[ \frac{1}{\beta} (T_n-T_j)^2 e^{-\beta(T_n-T_j)}+ \frac{2}{\beta^2} (T_n-T_j)e^{-\beta(T_n-T_j)}\right.\\
	 &&\hspace{1cm} \left.+ \frac{2}{\beta^3} \left(e^{-\beta(T_n-T_j)}-1\right)\right]\\
 	 &&+  \sum_{i=j}^{n}
	 \left[\frac{\gamma C(i)}{ \lambda+\gamma A(i) } -\left(\frac{\gamma B(i)}{\lambda+\gamma A(i)}\right)^2\right];
\end{eqnarray*}

\begin{eqnarray*}
\frac{\partial^2 \log L}{\partial \lambda^2}	 &=&-	\sum_{j=1}^{n} \left[\frac{1}{\left( \lambda+\gamma A(i)\right)^2 }  \right];\\	\frac{\partial^2 \log L}{\partial \beta \partial \gamma}	 &=& -
	\sum_{j=1}^{n} \left[ \frac{1}{\beta} (T_n-T_j)e^{-\beta(T_n-T_j)}+ \frac{1}{\beta^2} \left(e^{-\beta(T_n-T_j)}-1\right)\right]\\
	&&+  \sum_{i=j}^{n}
	\left[\frac{\gamma A(i)B(i)}{\left(\lambda+\gamma A(i)\right)^2} -\frac{B(i)}{\lambda+\gamma A(i)}\right];
	\\
		\frac{\partial^2 \log L}{\partial \gamma\partial\lambda}	 &=&
		 - \sum_{i=j}^{n}
		\left[\frac{  A(i) }{\left(\lambda+\gamma A(i)\right)^2} \right] ;
	\\
		\frac{\partial^2 \log L}{\partial \beta\partial\gamma}	 &=&  \sum_{i=j}^{n}
		\left[\frac{ \gamma B(i) }{\left(\lambda+\gamma A(i)\right)^2} \right].
\end{eqnarray*}
In the previous  equation the functions $B$ and $C$ are defined as $B(1)=0, B(j)=\sum_{i=1}^{j-1} (T_j-T_i) e^{-\beta(T_j-T_i)}, j\ge 2$ and  $C(1)=0, C(j)=\sum_{i=1}^{j-1} (T_j-T_i)^2 e^{-\beta(T_j-T_i)}, $  $j\ge 2$.
Since the log-likelihood is non-linear with respect to the parameters, the maximization if performed by using nonlinear optimazation tecniques \cite{Ozaki}.
 
\begin{table}[h]
	\centering
	{ \small
		\begin{tabular}{|l||c|c|c| }
			\hline
			\textbf{Case} & $\hat{\lambda}_0$ & $q_{5\%}(\hat{\lambda}_0)$ & $q_{95\%}(\hat{\lambda}_0)$  \\
			\hline
			\hline
			b.1  $\lambda_0=0.1$ & \textbf{0.0101} & 0.0061 & 0.0141  \\	
			b.2  $\lambda_0=0.1$  & \textbf{0.0105}  & 0.0059  & 0.0162  \\	
			b.3  $\lambda_0=0.1$  & \textbf{0.0095}  & 0.0057 & 0.0139 \\ 			
			b.4  $\lambda_0=0.1$  & \textbf{0.0088}  &0.0053 & 0.0126 \\
			\hline
			\textbf{Case}  & $\hat{\gamma}$ &$q_{5\%}(\hat{\gamma})$ & $q_{95\%}(\hat{\gamma})$    \\
			\hline
			\hline
			b.1  $\gamma=0$ &    \textbf{0.0046} & 3.84$\cdot 10^{-9}$ & 0.0267   \\
			b.2  $\gamma=0.05$  &     \textbf{0.0366} & 0.0146  & 0.0606 \\	
			b.3  $\gamma=0.15$  &       \textbf{0.0840} & 0.0472 & 0.1217 \\	
			b.4  $\gamma=0.3$  &     \textbf{0.1163}  & 0.0479 & 0.1850  \\
			\hline
			\textbf{Case}    & $\hat{\beta}$ &  $q_{5\%}(\hat{\beta})$ & $q_{95\%}(\hat{\beta})$  \\
			\hline
			\hline
			b.1  $\beta=0$ &    \textbf{0.6097} & 0.0407 & 0.8025     \\
			b.2  $\beta=0.08$  &     \textbf{0.0814} & 0.0318 & 0.1379  \\	
			b.3  $\beta=0.2$  &     \textbf{0.1414} & 0.0846 & 0.2113 \\
			b.4  $\beta=0.5$  &      \textbf{0.2782}& 0.1393 & 0.4421 \\
			\hline
		\end{tabular}
	}
	\caption{Estimated parameters of the component $X_2$, given $M=20000$ realizations of the component $X_2$ itself for each of the parameters set b.1 --b.4. }.
\end{table}
 
 We see that the estimated values are below the true values, especially for big values of $\gamma,\beta$. For the set of parameters (a), the value of $\beta$ is largely overestimated, but this is not a problem since the corresponding value of $\gamma$ are very small, and thus there is no observable self-excitement. 
 
 \medskip

Regarding the jump magnitude distribution, we fit the data via an MLE procedure by considering a  Generalized Extreme Value (GEV) distribution. It is a continuous distribution which may seen as  the approximation of the maxima of sequences of of independent and identically distributed random variables. It depends upon three parameters which allow to fit properly  the data.

\medskip

Finally, the estimation of the mean-reverting parameter $\alpha_2$ of the
jump component $X_2$ can be done by using the estimator defined in
\cite{Meyer-Brandis_2010}. Given a dataset $Y_2$ which we aim to
model with our jump process $X_2$, a consistent estimator for the
mean-reversion parameter $\alpha_2$ is
\begin{equation}
\label{eq: estimator alpha 2}
\hat{\alpha}_2=\log\Big(  \max_{1\le j \le N} \frac{Y_2(j-1)}{Y_2(j)} \Big).
\end{equation}
We will need an approximation of this estimator in our
estimation process, which is very closely related to one made
in \cite{Meyer-Brandis_2010}. The details will be discussed during
the data filtering process in the next section.

\section{A study case:  Italian electricity spot prices }

Here we discuss the more strongly computational part of the work, by starting from the description of the time series we take into consideration, that is 8 year of daily recorded Italian electricity spot prices. We first perform an explorative analysis on the time series, and after that, we discuss the problem of data filtering that we need to solve in order to obtain from rough data the different components of our model. In the end, we perform out-of-sample simulationsin order to predict future prices of electricity, discussing the results with some evaluation metrics like Winkler score and Pinball loss function.

\subsection{The time series}

\begin{figure}[h]
	\centering
	\includegraphics[width=0.75\textwidth]{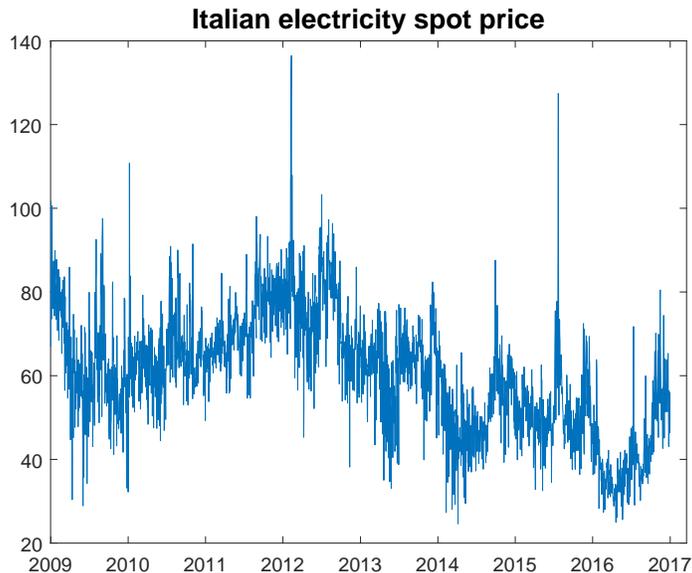}
	\caption{The time series $Y(t)$ of the the daily electricity prices
		in the MGP in the period 2009, January 1- 2017, December 31 ($t=1,\ldots,3287$). }
	\label{fig: rough data plot}
\end{figure}

We  consider the time series of the Italian
\textit{Mercato del giorno prima} (MGP, the day-ahead market),
available at \cite{MPG}. Figure \ref{fig: rough data plot} shows a
plot of the daily price time series $\{Y(t)\}_{t=1,\ldots,3287}$  from January,
$1^\mathrm{st}$ 2009 ($t=1$) to December, $31^\mathrm{st}$ 2017
($t=3287$).

\medskip

The MGP market is a day-ahead market, i.e.\ a market in which the
price is established via an auction for the blocks of electricity
to be delivered the following day. The agents that operate as
buyers in the market have to submit their offers between 8:00
a.m.\ and 12:00 noon of the previous day. The offers regard the
hourly blocks of electricity which will be delivered the following
day. This means that an agent will submit 24 different offers
(with different prices and quantities) for the electricity of the
following day, and he will do it all at the same time. Also the
sellers submit their offers, by telling the quantity of energy
that they are willing to sell and the price at which they want to
sell it. The \textit{market price} is then established before
12:55 p.m., and it is an hourly price, determined by finding the
intersection of the demand and the offer curve relative to the
specific hour of the day. After the determination of the market
price, all the electricity bought and sold for that hour is traded
at the market price.

\medskip

We choose to model the daily average of the hourly price.
This is a quite common choice in the literature, especially for
reduced-form models \cite{weron_review_2014}. Indeed, the main scope
of reduced--form models is to be able to capture more the medium
term (days/weeks) distribution of prices  than the hourly price,
and to use these estimated distributions of prices to pricing
electricity future contracts, which are very useful and used
financial instruments in the electricity market.

\medskip

We are aware that this averaging filters out many extreme
behaviours of the market. Indeed, a single hourly extreme price is
unlikely to produce a significant variation in the average daily
price. Anyway, we are also aware that reduced--form models usually perform
poorly on the hourly scale \cite{weron_review_2014}. We remark
that all the analysis that follows has been carried out also using
on--peak (08:00--20:00) and off--peak (20:00--08:00) data
separately, without obtaining a significant difference from the
entire day averages.

\medskip

The data available start from April, 2004, which is the moment in
which the liberalised electricity market started in Italy, but we
chose to focus on more recent data, from 2009 onwards, to make the
model more tight to the present nature of the electricity market.
This does not prevent the performance evaluation of the model from
being sufficiently robust, since the dimension of the sample is
$N=3287$.

\medskip

\begin{figure}[h!]
	\centering
	\includegraphics[width=0.8\textwidth]{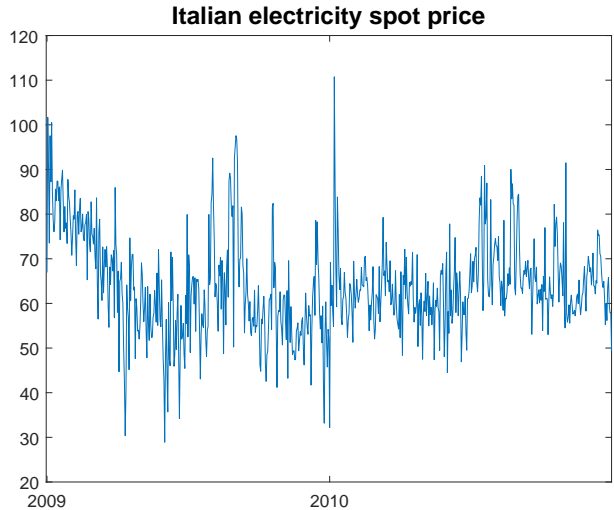}
	\caption{ The daily electricity prices
		in the MGP whitin the calibration window in the period  2009, January 1- 2010, December 31 ($t=1,\ldots,730$).}
	\label{fig: rough data plot_2009_2010}
\end{figure}

We use the data in the following way: the first $730$
days have been used for the study of the dataset and for the
validation of the model. Then, we evaluated the performance on
forecasting future prices for time horizons of length
$h=1,\dots,30$, using \textit{rolling windows}: at time $t$, we
use the data from $t-730$ to $t$ for the calibration of the
parameters of the model, and we simulate the future price at time
$t+h$ using those parameters. Then, we move ahead from time $t$ to
$t+h$ and we repeat the previous steps, starting from parameter
estimation.

\medskip

The first task  that has to be achieved 
is the separation, or \textit{filtering}, of the different signals in  the price time series $Y$. It is clear that this is not an easy task and it might not
be done in a unique way. In literature a lot of effort has been
done for this purpose (see
\cite{Janczura_spike_season_2013,Nowotarski_season_wavelets_2013,weron_review_2014,Meyer-Brandis_Tankov_2008,Meyer-Brandis_2010,Weron_2018}). Moreover, we think that in the present case the relatively small
presence of clearly recognisable spikes in the dataset makes the
spike identification task even more difficult than usual.

\medskip

Note that from now on we consider as the window for the
calibration of the model the one corresponding to the first two
years, 2009 and 2010. The reduced time series $Y(t), t\in [1,730]$ is
shown in Figure \ref{fig: rough data plot_2009_2010}.

\subsection{Data filtering}

\noindent{\bf Weekly seasonal component.} The first component that should be identified
and estimated is the one  dealing with trends and seasonality in
the data. As stated in \cite{Janczura_spike_season_2013} and in
the literature therein, the estimation routines  are usually quite
sensitive to extreme observations, i.e.\ the electricity price
spikes. Hence, one should first filter out the spikes, that often
are identified by the outliers. Actually, whether to filter out
the spikes before or after the identification of the deterministic
trends is still an open question in general.
Furthermore, the deseasonalizing methods used in literature are
very different: some authors suggest to considered sums of
sinusoidal functions with different intensities
\cite{cartea_figueroa_2005,geman_roncoroni_2006,Meyer-Brandis_2010},
others consider piecewise constant functions (or dummies)  for the
month \cite{fleten_2010,schwartz_2002}, or the day
\cite{de_jong_2006} or remove the weekly periodicity by
subtracting the \textit{average week}
\cite{Janczura_spike_season_2013}.
It turns out than an interesting and more robust technique is the method of wavelet
decomposition and smoothing,  applied among others in
\cite{Janczura_spike_season_2013,Nowotarski_season_wavelets_2013,weron_2006,weron_2008}.
\begin{figure}[h!]
	\centering
	\includegraphics[width=0.9\textwidth]{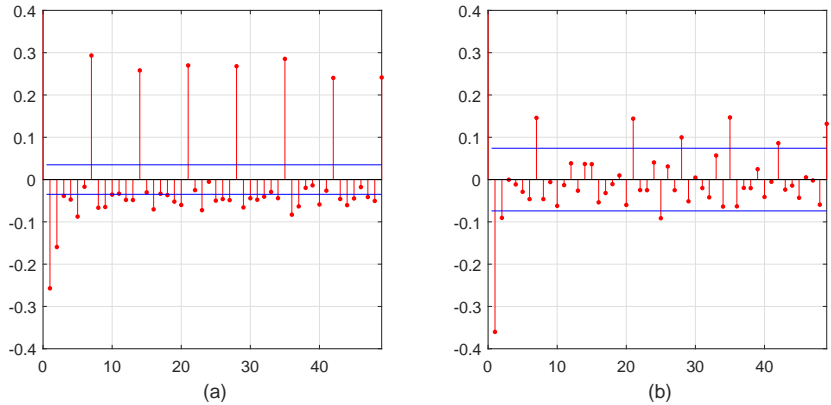}
	\caption{Autocorrelation function of the daily  electricity price returns. (a)  Full series 2009, January 1- 2017, December 31. (b)  Prices within the  calibration window  2009, January 1- 2010, December 31.}
	\label{fig:autocorrelation_whole_and_calibration_window}
\end{figure}

\medskip

Figure \ref{fig:autocorrelation_whole_and_calibration_window} shows the
autocorrelation function both for the whole series and within  the
calibration window. In the first case,   for any lag multiple
of seven the correlation is statistically significant, while in the
calibration windows only for the lag=14 there is not a significant
correlation at the level $95\%$. In any case the presence of
weekly periodicity is clear.
As a consequence we first filter the weekly
periodicity.  As in \cite{de_jong_2006,weron_review_2014} we do it
by means of   dummy variables, which take   constant value for
each different  weekday. Hence, we subtract the \textit{average
	week} calculated as the sample mean of the sub-samples of the
prices corresponding to each day of the week, as in
\cite{Janczura_spike_season_2013}. Public holidays  are treated in
this study as the eighth day of the week. Hence, the total number
of dummies is eight.

\medskip

More precisely, we define a function $D=D(t)$ which
determines which label the $t$-th day has, i.e.  $D(t)=i,
i=1,,\ldots,7,8$, if day $t$ correspond to Monday (and not a
festivity), ..., Sunday (and not a festivity) and a festivity,
respectively. We define the restricted time series $Y_{D=j}$ as the time series
formed only by the price values labelled with the day $j$, and
with $\overline{Y_{D=j}}$ its arithmetic mean. Then we define the
whole dummy variables function $Y_D$ as
\begin{equation}\label{def:dummies}
Y_D(t)=\sum_{j=1}^{8}1_{D(t)=j}(t)\overline{Y_{D=j}}.
\end{equation}
\begin{figure}[h!]
	\centering
	\includegraphics[width=\textwidth]{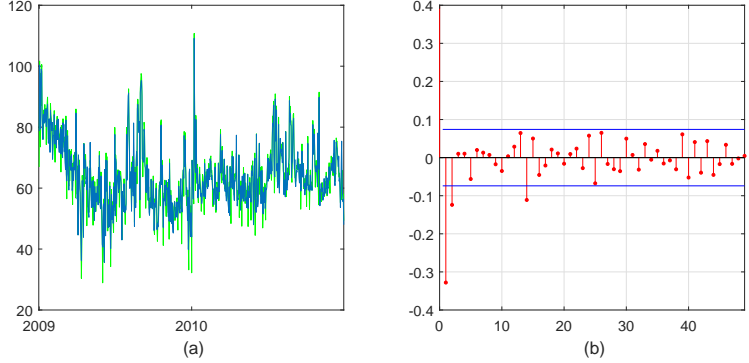}
	\caption{(a) Original prices $Y$ (green) and prices after dummies removal $Y_w$ (blue);
		(b) Autocorrelation of the     returns of the prices $Y_w$}
	\label{fig: data plot after dummies}
\end{figure}

The time series obtained after the weeekly deseasonalization   is
the following
\begin{equation}
Y_{w}(t):=Y(t)- \hat{f}_s(t)=: Y(t)-(Y_D(t)-\overline{Y}), \label{def:price_no_dummies}
\end{equation}
where   $\overline{Y}$ is the arithmetic mean of $Y(t),
t=1,...,730.$  The arithmetic mean of the prices $Y_w$ corresponding to a specific
day of the week coincides with the mean of   all prices within
the calibration window.  Figure \ref{fig: data
	plot after dummies}-(a) shows that  $Y_w$ after removing
the dummy variables is barely distinguishable from the original
time series $Y$; it is only a bit more regular. On the other hand,
$Y_w$ does not show weekly correlation, as visible in Figure
\ref{fig: data plot after dummies}-(b). The function $\hat{f}_s$ is the estimate of the short seasonality in $f$.

\medskip

\noindent\textbf{Jump component.} Before filtering the long-term seasonal component, we filter out the jump component. The reason for this order lies in the following consideration: at a small scale, the presence of a slowly moving seasonal trend does not affect the recognition of a price spike. On the other hand, if we chose to filter the long-term seasonal component before filtering out the spikes, the presence of one of more spikes could affect the form of the seasonal component, which is something that we intuitively regard as incorrect. Indeed, we tend to consider the spikes as an "external event", and we do not want the seasonal term to be affected by the presence of a price spike.

\smallskip

 As in \cite{Meyer-Brandis_2010}, the idea is to obtain the filtered time series,   denoted   $Y_J$, as the series that contains the jumps and their paths of reversion towards their mean. We first   estimate the mean-reversion speed $\alpha_2$  by   the estimator $\hat \alpha_2$ given by \eqref{eq: estimator alpha 2}; by performing the estimate along   the entire time series, not only in the jump times: this is not restrictive, since the strongest rates of reversion towards the mean happen right after a jump has occurred. Afterwards we identify the  jump times.  The idea is to consider as jumps the price increments that exceed $k$ standard deviations of the price increments time series. This cannot be implemented to the time series $Y_w$ directly, because, in case two spikes appear one after the other, the second one would not be considered as a spike. In order to avoid this effect, we define the modified time series $\tilde{Y}_w$ as
$$\widetilde{Y}_w(t):=(1-\alpha_2)Y_w(t)+\hat \alpha_2 \bar{Y}_{30}(t),$$
where $\bar{Y}_{30}(t)$ is the moving average of the time series $Y_w$ over periods of 30 days. Then, we defined the times series 
\begin{equation}\label{eq:Y_MOD}
\left\{ Y_w(t)-\widetilde Y_w(t-1)  \right\}-{t=2,3,\ldots}
\end{equation}
 of the modified increments, which takes into account a reversion effect towards the moving average $Y_{30}$. It  performs very well also when the spikes appear in clusters. 
 Then,  denoted by $\widetilde{\sigma}$ the standard deviation of the series \eqref{eq:Y_MOD}, we say that a spike occurs at time $\tau$ if
 \begin{equation}\label{eq:jump_rule}
 \left| Y_w(\tau)-\widetilde Y_w(\tau-1)\right| > 2.5 \widetilde{\sigma}.
\end{equation}

If $N$ is the number of detected spikes, i.s. if $\left\{\tau_j \right\}_{j=1,\ldots,N}$ are the times satisfying condition \eqref{eq:jump_rule}, the corresponding jumps are defined, for $j=1,\ldots,N$ as $$\hat{\mu}_j=Y_w(\tau_j)-\widetilde Y_w(\tau_j-1).$$

\medskip

Once we have the estimates $\left\{(\tau_j,\hat{\mu}_j) \right\}_{j=1,\ldots,N}$ of the times and magnitudes \eqref{eq:(T,Z)}
of the jumps, we obtain the estimation $Y_J(t)$ of    the solution $X_2$ of \eqref{eq:X_2_jump} as follows 
\begin{equation}
\label{eq:Y_J}
Y_J(t) = \sum_{j=1}^N  \mu_j e^{-\hat\alpha_2(t-\tau_j)} \epsilon_{\tau_j}([-\infty,t])
\end{equation}

\medskip

Given $Y_J$ as in \eqref{eq:Y_J}, we denote the filtered time series as
$$Y_s(t)=Y_w(t)-Y_J(t).$$

\medskip

\noindent\textbf{Long-term seasonal component.} Now we look for an  estimator $\hat{f}_l$ of the long-term
seasonality component, so that 
$$\hat{f}= \hat{f}_s + \hat{f}_l,$$
where $\hat{f}_s$ is given in \eqref{def:price_no_dummies}, is an estimator of the   deterministic part $f$ in \eqref{eq: model basic}.
 There is a lot of literature on the subject (see, for example, 
\cite{benth_Kiesel_Nazarova_2012,cartea_figueroa_2005,de_jong_2006,geman_roncoroni_2006,Janczura_spike_season_2013,schwartz_2002,weron_2008}). These references
try to explain such a component by means of sinusoidal
functions or sums of sinusoidal functions of different
frequencies. In our case  it seems
that there is no statistically significant dependence upon such
periodic function, both in the case of monthly, half-yearly or yearly
periodicity.

\medskip

We apply  the method of wavelet
decomposition and smoothing,  applied among others in
\cite{Janczura_spike_season_2013,Nowotarski_season_wavelets_2013,weron_2006,weron_2008}. The idea is to consider   repeatedly the  convolution of   $Y_s$   with a family of wavelets, which have the effect of smoothing out    $Y_s$. If we manage to smooth $Y_s$ enough to remove the effect of stochastic oscillation, but not too much to remove the long-term trend, then we can subtract this smoothed version of $Y_s$ from $Y_s$ itself, obtaining a centred time series with almost no long-term oscillation.

\begin{figure}[h!]
	\centering
	\includegraphics[width=0.8\textwidth]{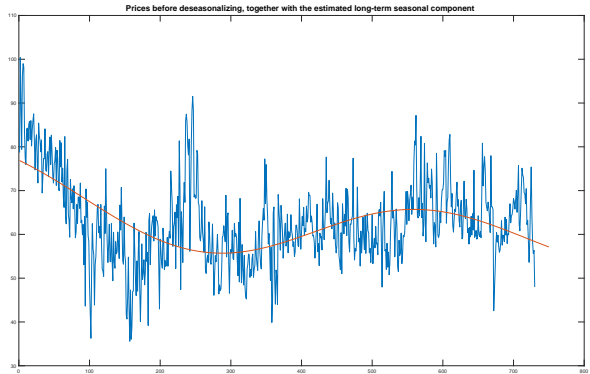}
	\caption{Price series $Y_S$ (blue) and estimation function $\hat{f}_l$ (red) via Daubechies of order 24   wavelets at level 8.}
	\label{fig:wavelets_and_Y_1}
\end{figure}

We go more into details: we refer to 
\cite{Nowotarski_season_wavelets_2013} and the literature therein. 
We use wavelets belonging to the Daubechies family, of order
24, denoted by (F--db24). Wavelets of different families and orders make
different trade-offs between how compactly they are localized in
time and their smoothness. Any function or signal (here,
$Y_s$) can be built up as a sequence of projections onto
one father ($W$) wavelet and a sequence of mother wavelets ($D$):
$Y_s=W_k + D_k + D_{k-1} + ... + D_1$, where $2^k$ is the maximum
scale sustainable by the number of observations. At the coarsest
scale the signal can be estimated by $W_k$. At a higher level of
refinement the signal can be approximated by $W_{k-1} =W_k + D_k$.
At each step, by adding a mother wavelet $D_j$ of a lower scale $j
= k - 1, k - 2, ...,$ one obtain a better estimate of the original
signal. Here we use $k= 8$, which corresponds a quasi-annual ($2^8
= 256$ days) smoothing.
Then the estimator $\hat{f}_l$ of the long-term deterministic part $f-\hat{f}_s$ is given by
\begin{equation}\label{eq:hat_f_l}
\hat{f}_l(t)=W_8(t),
\end{equation}
the
Daubechies wavelets of order 24 at level 8.

\medskip

The resulting  time series $Y_f$ 
given by
\begin{equation}
Y_f(t)=Y_s(t)-\hat{f}_l(t),\label{eq:Y_1}
\end{equation}
represents a realization of the base component $X_1$.

\medskip

Figure \ref{fig:wavelets_and_Y_1} show the price series $Y_s$ and
the overlapped estimated $\hat{f}_l$.   
The wavelet interpolation is here extended outside
the calibration window. This is not automatic in the case of wavelet decomposition, since the wavelets
are compactly supported and we are convolving only up to the final time of our dataset. To obtain this prolongation, we prolonged the time series in the forecasting window by using the technique of exponential reversion to the median,
thoroughly studied in \cite{Nowotarski_season_wavelets_2013}, before applying the wavelet de-noising.
In this way we have been able to obtain a function $\hat{f}_l$ which extends also to times $t$ in the forecasting windows.

\subsection{Out of sample simulations}

We will perform and assess here the forecasts of future electricity prices through our model. We first outline the simulation scheme and make some considerations about the parameters in the rolling windows. After this, we define the metrics which we will subsequently use to evaluate our results.

\subsubsection{Parameter estimation in the rolling windows}

We recall that for each time $t$, when forecasting the price
distribution at time $t+h$ (where $h\in\{1,\dots,30\}$ is the
forecasting horizon), we carry out a new calibration of all the
parameters of the model, including the Hurst coefficient $H$. We
think that in this way, if there is a change in the data coming as
input, the model is able to change its fine structure coherently
with these changes. For example, if the self--correlations
changes, or disappears, at some point, we expect the parameter $H$
to change consequently, and possibly approach $\frac 12$.

\smallskip

\begin{figure}[h]
	\includegraphics[width=0.9\textwidth]{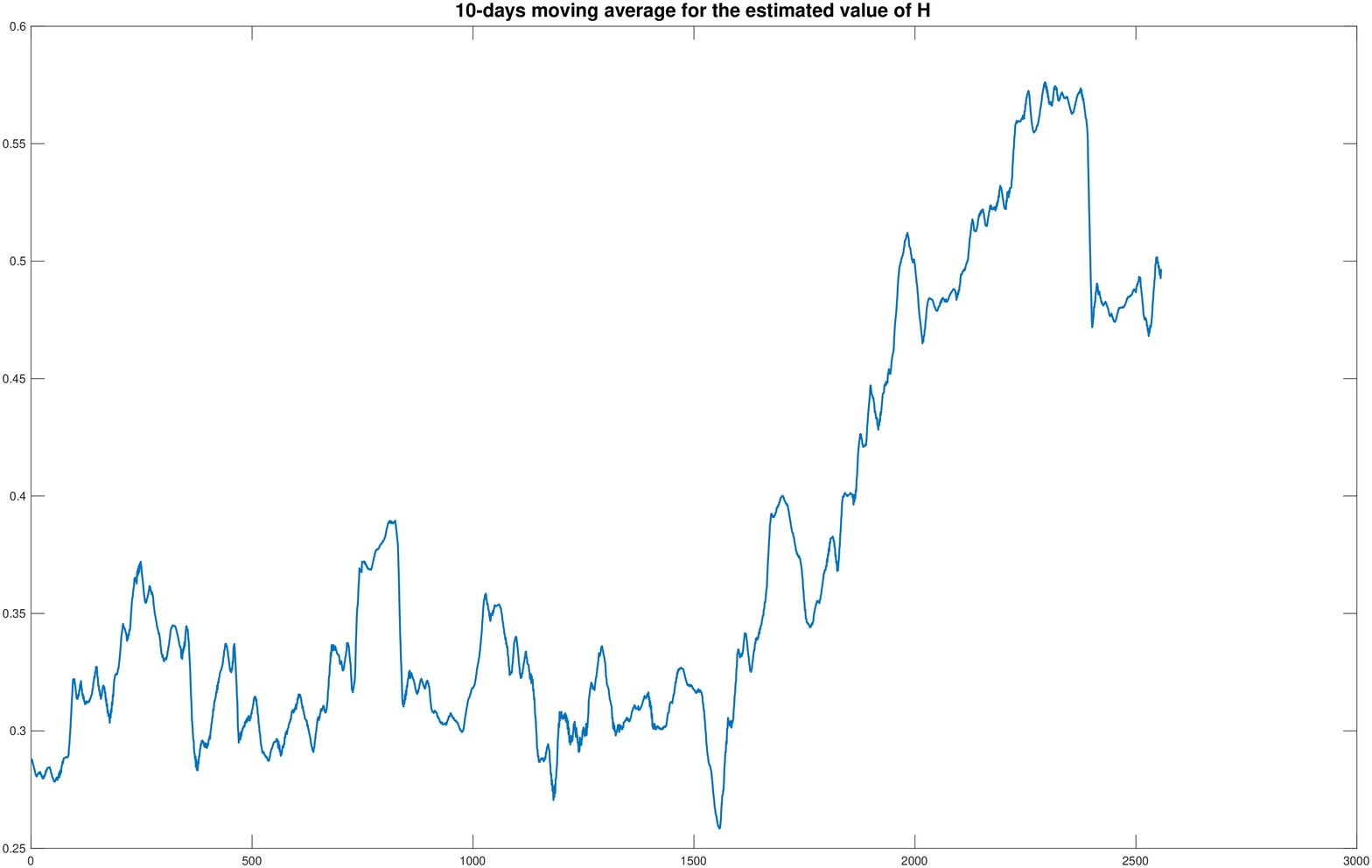}
	\caption{\small{10-day average of the estimated value $\hat{H}$ of $H$. We chose the 10-day average in order to smooth out some irregular behaviour at shorter scale. }}
	\label{fig: H}
\end{figure}

We start by analysing $H$: as long as our rolling window for
estimation is advancing,the values of $H$ are on average
increasing. The first estimate is
$\hat{H}=0.2909$. The mean value across the whole dataset is
$0.3735$. In general, the values are such that
$$0.2128\leq \hat{H}\leq 0.6234.$$
Notice that the maximum value is above the $H=1/2$ threshold and
this shows that a positive correlation between the increments
may occur.  Figure \ref{fig: H} shows the pver ten days averaged behaviour of $H$ across all    time lengths in.

\medskip

In general, as we already pointed out,
we think that this moving identification of the parameter $H$ is
useful to update the fractional structure of the model when the
input data are suggesting to do so, giving a better modelling
flexibility, also in case of future changes in the market nature.

\medskip

Regarding the parameters of the fractional Ornstein--Uhlenbeck
process $X_1$, $\alpha_1$ and $\sigma$, progressing in our rolling
window, we found a change in the estimated values of the
parameters; summarizing, we obtained that the range of values for
$\hat{\alpha}_1$ and $\hat{\sigma}$
\begin{equation*}
\begin{split}
& 0.0453 \leq \hat{\alpha}_1 \leq 0.6823 \\
& 4.2069 \leq \hat{\sigma} \leq  8.8972.
\end{split}
\end{equation*}

We see that there is a large variation in the parameters,
especially for $\hat{\alpha}_1$, but we remark that this variation
is gradual, since the mean reversion rate decreases while moving
through our forecasting dataset, together with the volatility
$\hat{\sigma}$. Looking at Figure \ref{fig: rough data plot}, this
can be observed, at least for the volatility, also by a
macroscopic observation of the data.

\medskip

\begin{figure}[h]
	\includegraphics[width=0.9\textwidth]{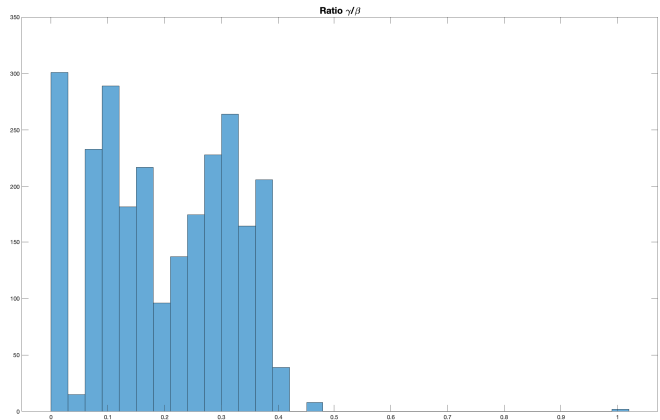}
	\caption{\small{Ratio between $\hat{\gamma}$ and $\hat{\beta}$ throughout the entire dataset. }}
	\label{histogram ratio}
\end{figure}

Looking at the estimation of the parameters of the jump process $X_2$,  a bit of variability   through the dataset is shown, but there is not an evidence of a particular pattern. Moreover,   the jump observations are relatively rare (20-40), so  based on the available data, it would be even more difficult to draw conclusions on their long term behaviour.	
The estimated values of the mean--reversion
coefficient $\alpha_{2}$ are suche that $ \hat{\alpha}_2 \in [0.3564, 0.6211]$, with an the mean value   $0.4358$. As the Hawkes process parameter estimation regards, $\hat{\lambda}_0 \in[ 0.0101 , 0.0284]$ with mean value   $0.0169$,   $\hat{\gamma} \in [1.12\cdot 10^{-9} , 0.1574 ]$, with mean their mean $0.0625$ and  $\hat{\beta} \in [ 0.0012 , 0.9993]$ with mean $0.3662$. 
There is a great variation in such estimates. This, in our opinion, is due to the low dimension of the dataset, because not many spikes are present. Thus, the MLE method is finding sometimes a good evidence of a self-excitement (when $\gamma$ is bigger and relatively close to $\beta$) and sometimes no evidence of self-excitement (when $\gamma$ is very small and/or $\beta$ is much bigger than $\gamma$).  To show this fact, we plot in Figure \ref{histogram ratio} an histogram of the ration $\gamma/\beta$, which is a very good indicator of the presence of self-excitement.  From the histogram, we can see that roughly half of the time the ratio is below 0.25 (showing little or no self-excitement), and half of the time above (showing a significant self-excitement). We think this is another evidence of the flexibility of our model, similarly to the estimations of $H$. If some self-excitement seems to be present, then the model is including it. Otherwise, the model will simply produce a classical point process with constant intensity.

\begin{figure}[h]
	\includegraphics[width=0.85\textwidth]{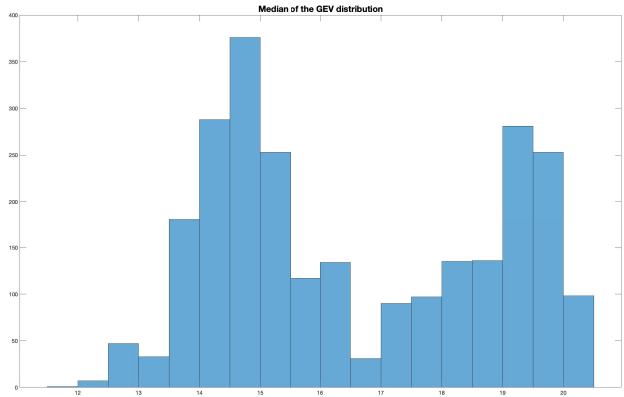}
	\caption{\small{Median of the GEV distribution throughout the entire dataset. }}
	\label{median GEV}
\end{figure}

As the  estimation of the parameters  $(\tilde{\mu}, \xi,\tilde{\sigma})$ of the Generalized Extreme Value distribution  we have that $  \hat{\tilde{\mu}} \in [11.2506, 19.4064]$ with a mean value  $15.6687$, $\hat{\xi} \in [
-0.4809, 4.2591]$ with mean  $0.4125$  and $\hat{\tilde{\sigma}} \in [0.3248, 3.8260]$ with mean $2.2331$.  We only remark that  even if there is a great variability in the parameters, the median value of the jump size is not varying from one estimate to the other. We recall that the median (the mean is not always defined) of a GEV distribution is given by
$$\text{Median}=\mu+\sigma\frac{\log(2)^{-\xi}-1}{\xi}.$$
In Figure \ref{median GEV}, we see that this value is oscillating between 11 and 22, which are reasonable values for our dataset.

\subsubsection{Forecasting performance}
\label{sec: forecasting performance}
Now we evaluate the performance of the model described in Section \ref{sec: model proposal} for the forecast of future
values of the electricity prices. As pointed out in \cite{weron_review_2014}, there is no universal standard procedures for evaluating the forecasts.

\medskip

The most widely used technique is to obtain point forecasts, i.e.\ single forecast values, and evaluate them using some error function.
The most common error function for this type of forecasts is Mean Absolute Percentage Error (MAPE), together with its refinement Mean Absolute Scaled Error (MASE).
Another frequently used measure is the Root Mean Square Variation (RMSV), which is simply the estimated standard variation of the forecast error.
%
In our model, the SDE-type structure is not particularly suitable
for giving short-term point forecasts, as it is also pointed out
in \cite{weron_review_2014}. So in the following we will not
concentrate our analysis on point forecasts, as we do not expect
our model to be able to outperform more sophisticated and
parameter-rich model in this task.

\medskip We focus on the relatively novel concept of \emph{Interval Forecast}.
 More recently, as was already suggested in \cite{weron_review_2014} and as it has been more thoroughly analized in the very recent review
\cite{Weron_2018}, the driving interest in forecast evaluation has been put in interval forecasts and density forecasts. Interval forecast
have also been used as the official evaluating system in 2014's Global Energy Forecasting Competition (GEFCom2014). There is a close relation
between interval forecasts and density forecasts.

Interval forecasts (also called Prediction intervals, shortly PI) are a method for evaluating forecasts which consists in constructing the intervals
in which the actual price is going to lie with estimated probability $\alpha$, for $\alpha\in(0,1)$. There are many technique for the construction of the intervals and their evaluation  \cite{Weron_2018}. Here we consider interval forecasts for different time lags $h$, by means of  different techniques: Unconditional Coverage (UC),
Pinball loss function (PLF) and Winkler Score (WS).
\medskip

\noindent\textbf{Unconditional Coverage (UC)}. Establishing the UC just means that we evaluate nominal rate of coverage of the Prediction intervals;
as stated in \cite{Weron_2018}, one can simply evaluate this quantity, or consider the average deviation from the expected rate $\alpha$. As pointed out in \cite{Weron_2018}, if we call $P_t$ the actual price, $[\hat{L_t},\hat{U_t}]$ the Prediction
interval at level $\alpha\in (0,1)$, we are checking the fact that the random variable
$$I_t=\begin{cases}
1, & \text{if } P_t\in [\hat{L_t},\hat{U_t}],\\
0, & \text{if } P_t\notin [\hat{L_t},\hat{U_t}],\\
\end{cases}$$
has a Bernoulli $B(1,\alpha)$ distribution. This works clearly under the assumption that the violations are independent, which may not always be the case.

\medskip

\noindent\textbf{Pinball loss function (PLF)}. The Pinball loss function is   a scoring function which can be calculated for every quantile $q$. If we denote with $P_t$ the actual price,
with $Q_q(\hat{P_t})$ the $q$-th quantile of the estimated prices $\hat{P_t}$ obtained with the model, the Pinball loss function is defined as
\begin{equation*}
\begin{split}
\text{Pin}(Q_q(\hat{P_t}),P_t,q):=\begin{cases}
(1-q)(Q_q(\hat{P_t})-P_t), & \text{if } P_t<Q_q(\hat{P_t}),\\
q(P_t-Q_q(\hat{P_t})), & \text{if }P_t \geq Q_q(\hat{P_t}).
\end{cases}
\end{split}
\end{equation*}

\medskip

\noindent\textbf{Winkler Score (WS)}. The Winkler score is a scoring rule  similar to the Pinball loss function, with the aim of rewarding
both reliability (the property of having the right share of actual data inside the $\alpha$-th interval) and sharpness (having smaller intervals).
For a central $\alpha$-th interval $[\hat{L_t},\hat{U_t}]$, $\delta_t:=\hat{U_t}-\hat{L_t}$, and for a true price $P_t$, the WS is defined as
$$\text{WS}([\hat{L_t},\hat{U_t}],P_t)=\begin{cases}
\delta_t, & \text{if }P_t\in[\hat{L_t},\hat{U_t}] ,\\
\delta_t+\frac{2}{\alpha}(P_t-\hat{U_t}), & \text{if }P_t>\hat{U_t},\\
\delta_t+\frac{2}{\alpha}(\hat{L_t}-P_t), & \text{if }P_t<\hat{L_t}.\\
\end{cases}$$
As it can be seen, the WS has a fixed part which depends only on the dimension of the Prediction intervals.

\medskip

\medskip

\noindent\textbf{Comparison of different models}. We checked the performance not only of the model proposed in Section \ref{sec: model proposal}, but of two  more  models, in order to highlight the peculiarities of the model proposed in this paper. The examined models are the following.
\begin{itemize}
	\item[i)] The stochastic differential equations  described in Section \ref{sec: model proposal} in which the base component is driven by a fractional Brownian motion.
		\item[ii)] The same model as in i) in which instead of the fractional Brownian motion, the standard Brownian motion is considered.  
	\item[iii)]A naive model, built as $\text{Naive(t)=D(t)+H}$, where $D$ is the dummy variables function and $H$ is randomly sampled
	historical noise  coming from the relative calibration window \cite{Weron_2018}.
\end{itemize}

\medskip

\noindent\textbf{Forecasting horizons}. As already mentioned, used as the calibration window a rolling window with fixed dimension of $730$ prices,
corresponding to the $730$ days of past observations. In this framework, we will consider the following forecasting horizons $h$:
$$h=\{1,2,\dots,29,30\}$$
For each forecasting horizon we make a new estimate of the parameters at a distance $h$ from the previous one, in order to have the sampled
prices coming from disjoint time intervals.

\subsubsection{Results}

\begin{figure}[h!]
	\includegraphics[width=\textwidth]{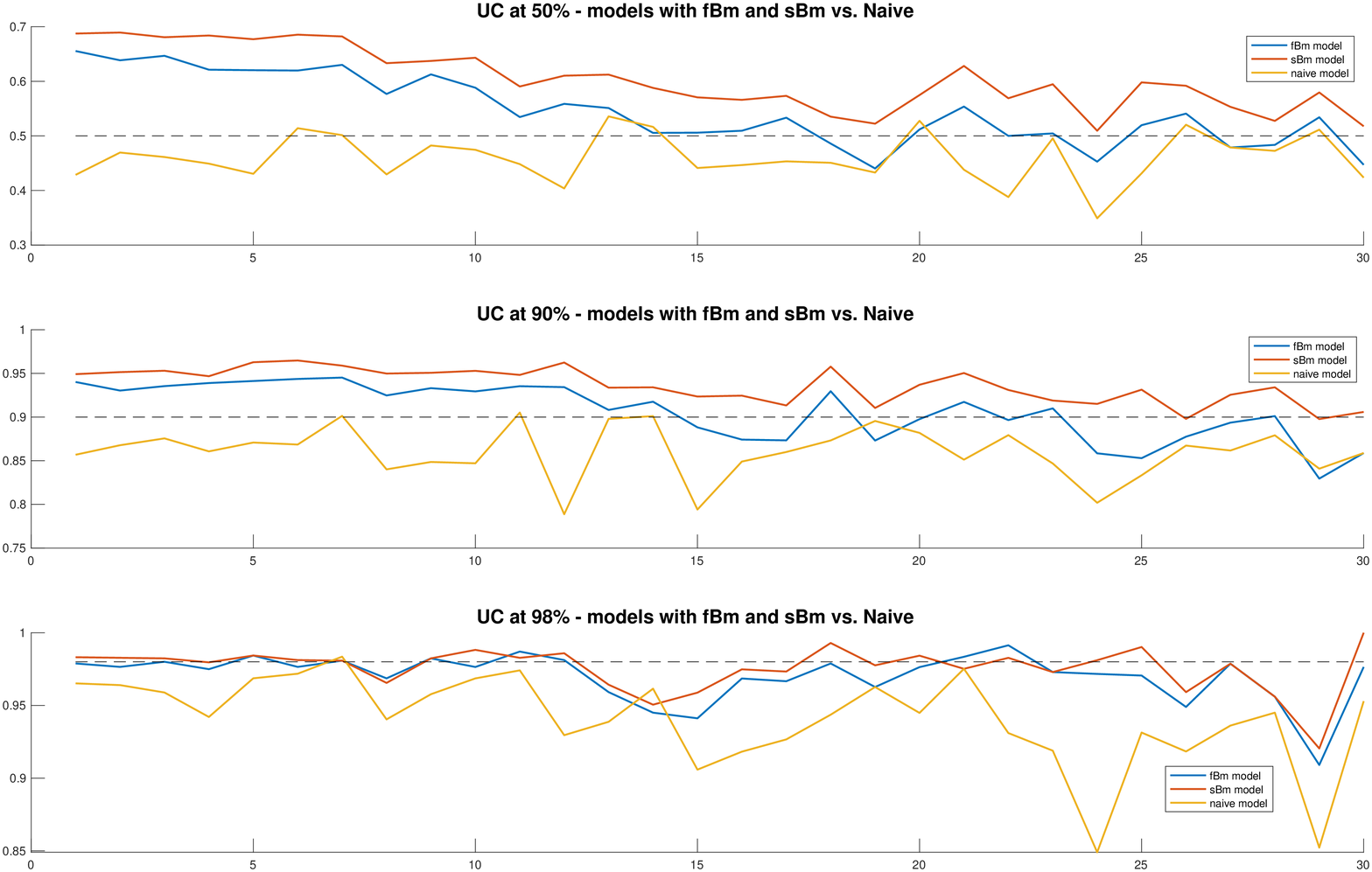}
	\caption{\small{Observed Unconditional Coverage of the model with the fractional Brownian motion (blue), with the standard Brownian motion (red) and the naive one (yellow))  for $50\%,90\%$ and $98\%$ coverage intervals. On the horizontal axis, we represent the length $h$ of the forecasting window we are considering, while on the vertical axis we represent the UC value.}}
	\label{fig: UC plot wavelets}
\end{figure}

We start by analysing the performance of the models by their observed Unconditional Coverage.  Figure \ref{fig: UC plot wavelets} shows their performance in a plot which spans across all the forecasting horizons $h$ that we are considering. The dotted black line represents the relative level of coverage that we should attain. The closer we are to the dotted line, the more accurate a model is in covering that specific interval.

In the $50\%$ interval (top plot), the na\"ive model seems to  be more stable, even if it is almost always under-covering the interval. Among our models, the fBm model, except for the shorter forecasting horizons, is performing remarkably well. The sBm model is over-covering the interval, for almost all forecasting horizons. 

Moving to the $90\%$ (mid plot) and $98\%$ (bottom plot) prediction intervals, the fBm model performs in general better than the other ones, except for a slight
excess in coverage for the $90\%$ PI with small forecasting windows. In the $98\%$ prediction intervals, also the sBm model has a very good performance, which will be confirmed by the numerical data for the Unconditional Coverage reported in Table \ref{tab: UC}.

\medskip

\begin{figure}[h]
	\includegraphics[width=0.85\textwidth]{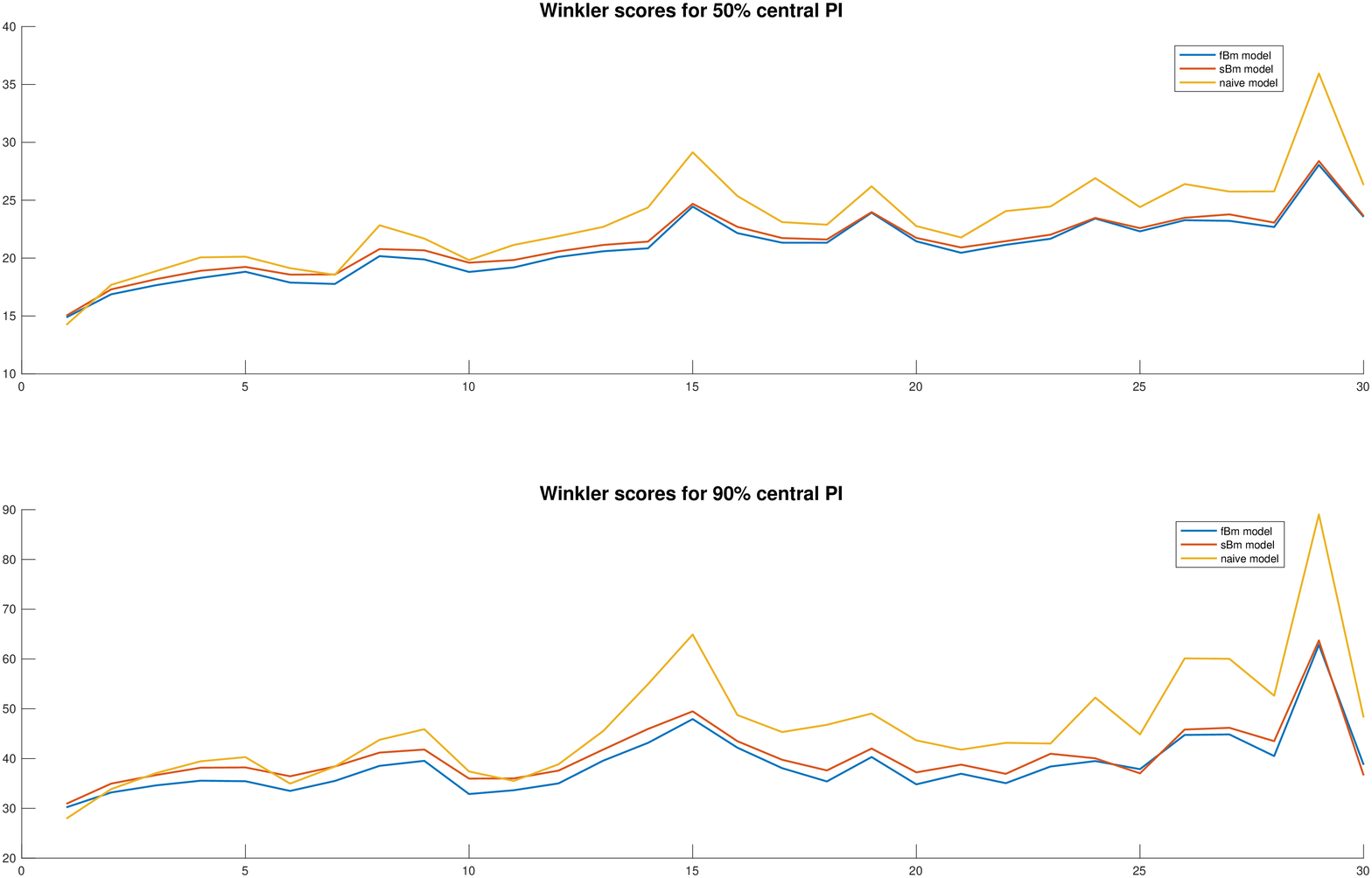}
	\caption{\small{Winkler scores for $50\%$ and $90\%$ PI. Again, on the horizontal axis, we represent the length $h$ of the forecasting window we are considering, while on the vertical axis we represent the Winkler score value.}}
	\label{fig: WS}
\end{figure}

\begin{figure}[h]
	
	\includegraphics[width=0.85\textwidth]{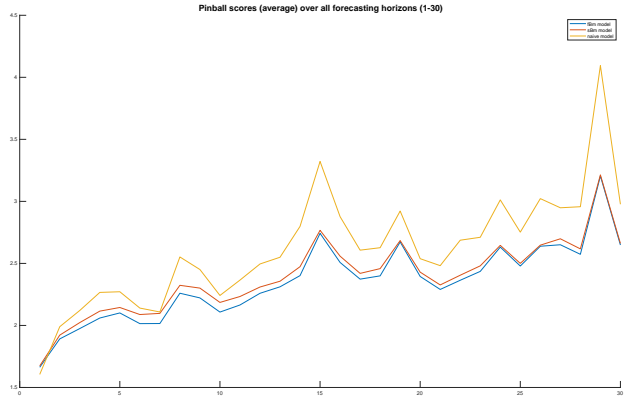}
	\caption{\small{Pinball loss function (average).On the horizontal axis, we represent the length $h$ of the forecasting window we are considering, while on the vertical axis we represent the Pinball loss function value.}}
	\label{fig: PLF}
\end{figure}

We analysed then the Pinball loss function and  Winkler Score   of the three models. In Figure \ref{fig: WS} and in Figure \ref{fig: PLF} we reported again the
results spanning along all forecasting horizons. We note that the Pinball loss function is a function of the quantile we are evaluating, so that in principle
we would have to evaluate it separately for every quantile $q=1,\dots,99$. As it has also been done in the GEFCom2014 competition, we averaged first over all
quantiles, in order to obtain a single value and make comparisons easier.

\medskip

In terms of the Winkler scores (Figure \ref{fig: WS}), the fBm  and the sBm models outperform the  the na\"ive benchmark. The difference between the fBm model and the sBm model is very small in general, but the fBm model performs better than the sBm one in almost every prediction interval. This is true especially if we consider the $90\%$ interval WS.

There seems to be a sort of contradiction in our results: indeed, when it comes to the UC at level $50\%$, the na\"ive model seemed to be slightly better than the fBm model, while in terms of WS for the $50\%$ PI the fBm is clearly superior to the na\"ive benchmark. This is possible because the WS is a metric which not only evaluates the share of coverage of a prediction interval, but also gives a penalty for missed values, and this penalty depends on the magnitude of the error made. Thus, it is seem reasonable to suppose that the na\"ive model, while performing quite good in terms of coverage at the $50\%$ PI, makes bigger errors than the fBm model.

The results about the Pinball loss function  are quite similar to the ones of the Winkler scores. Again, the fBm and the sBm model outperform the na\"ive model, while being very close one to each other. Again, the fBm model performs slightly better than the sBm model.

\medskip

Both in terms of WS and Pinball loss function, the na\"ive model is performing better than the fBm and the sBm model when the forecasting horizon $h=1$. This is somewhat consistent with a fact mentioned in \cite{weron_review_2014}, which we already reported: the reduced-form models, like ours, usually have a quite poor performance in very short-term forecasts.

\begin{table}[h]
	\centering
	{\small
		\begin{tabular}{|l||c|c|c|}
			\hline
			\textbf{Avg.\ score\textbackslash Model} &  fBm & sBm & Na\"ive  \\
			\hline
			\hline
			$UC_{50\%}$ & $54.54\%$ & $60.37\%$ &  $\mathbf{46.02\%}$  \\
			\hline
			$UC_{50\%}$ error  &$+4.54\%$ & $+10.37\%$ & $\mathbf{-3.98\%}$ \\
			\hline
			$UC_{50\%}$ abs.\ error & $5.95\%$  & $10.37\%$ & $\mathbf{4.83\%}$  \\
			\hline
			\hline
			$UC_{90\%}$ & $\mathbf{90.63\%}$ & $93.65\%$ & $86.02\%$ \\
			\hline
			$UC_{90\%}$ error & $\mathbf{-0.93\%}$ & $-9.10\%$  & $+3.79\%$\\
			\hline
			$UC_{90\%}$ abs.\ error & $\mathbf{2.81\%}$ & $3.67\%$ & $4.04\%$ \\
			\hline
			\hline
			$UC_{98\%}$ & $97.02\%$ & $\mathbf{97.58\%}$ & $94.13\%$  \\
			\hline
			$UC_{98\%}$ error & $-0.98\%$ & $\mathbf{-0.42\%}$ & $3.87\%$  \\
			\hline
			$UC_{98\%}$ abs.\ error& $1.19\%$ & $\mathbf{0.99\%}$ & $3.90\%$ \\
			\hline
		\end{tabular}
	}
	\caption{\small{Coverage rate for the estimated PI, averaged over all forecasting horizons $h=1,\dots,30$. The average error,
			as it can be seen, is just the difference of the average coverage from the nominal value.}}
	\label{tab: UC}
\end{table}

\begin{table}[h]
	\centering
	{\small
		\begin{tabular}{|l||c|c|c|}
			\hline
			Score \textbackslash Model & fBm & sBm & Na\"ive \\
			\hline
			\hline
			WS$_{50\%}$ & $\mathbf{20.87}$ & $21.30$ & $23.14$ \\
			\hline
			WS$_{90\%}$ & $\mathbf{38.62}$ & $40.44$ & $46.25$ \\
			\hline
			\hline
			PLF & $\mathbf{2.3484}$ & $2.3920$ & $2.6164$  \\
			\hline
		\end{tabular}
	}
	\caption{\small{Winkler scores and Pinball loss function values, averaged over all forecasting horizons $h=1,\dots,30$.}}
	\label{tab: WS and PLF}
\end{table}

In Table \ref{tab: UC} and Table \ref{tab: WS and PLF} we reported the above discussed values, averaged over all different forecasting horizons.
In terms of the UC, each of the three models is performing better than the others for a certain PI, while both for the WS and the PLF we see the better performance of the fBm model also from these numerical data.

\subsection{Conclusions}

Drawing some conclusions from the results analyzed above, there are some evidences that a fBm-driven model may be more
adequate to model the electricity prices than a sBm-driven model.

Regarding the forecasting performance (QF and PI), the fBm methods have
better performance than the sBm ones in terms of WS and PLF, while both the sBm and the na\"ive model enjoy some success when evaluating the UC.

To understand this apparent contradiction, we remark (as we already did) that WS and PLF are
scoring rules which give a penalty for missed forecasts (while UC does not),
and these penalties depend also on the magnitude of the error.
The fact that fBm models outperform sBm models in this evaluation
may mean that the QF and the PI given by the fBm models are in some
sense more robust than the sBm ones (and also than the ones of the
naive benchmark).

\medskip

Concerning the model structure, we remark that we found very satisfactory the fact that the parameter estimation for the Hawkes process gave roughly half of the times a very significant value, meaning that the clustering effect is not only visible on a macroscopic scale, but is also captured by the numerical methods. 

\medskip

This was not assured in principle, since the Italian
market  is rather peculiar, having only a small number of real spikes. This gives,
as a consequence, that the intensity of the spike process is small and could
become difficult to estimate, even if this was not the case for our data.

\medskip

Regarding the role of the fractional Broenian motion in the model, we remark that we found some very interesting informations from the estimation procedure. The fact, shown in Figure \ref{fig: H}, that the parameter $H$ is tending to $0.5$ in more recent times may mean that the market is finding automatically a way towards the "independence of increments", which would be implied by the fact that $H=0.5$. This is remarkable also for the fact that the independence of increments is closely related, for these models, with the absence of arbitrage. Even if we pointed out that arbitrage is usually not possible, for this kind of models, when trading only once per day, a good question for future developments in this sense may be: are electricity markets, which are "young" financial markets, finding their own stability with the passing of time, or are our findings specific to the italian market? In any case, are these changes going to last in the future or we may see a return of a fractional effect in the next years?

\end{document}